\documentclass[superscriptaddress,nofootinbib,a4paper,11pt]{article}
\usepackage{jheppub}
\usepackage{graphicx}
\usepackage{subcaption}
\usepackage{float}
\usepackage{amsmath}
\usepackage{amssymb}
\usepackage[utf8]{inputenc}
\usepackage{hyperref}
\usepackage{color}
\usepackage{slashed}
\allowdisplaybreaks

\newcommand{\be}{\begin{equation}}
\newcommand{\ee}{\end{equation}}

\begin{document}

\title{\boldmath The quirk signal at FASER and FASER 2}

\author[a]{Jinmian Li,}
\author[b,c]{Junle Pei,}
\author[a]{Longjie Ran,}
\author[d]{and Wenxing Zhang}

\affiliation[a]{College of Physics, Sichuan University, Chengdu 610065, China}
\affiliation[b]{CAS Key Laboratory of Theoretical Physics, Institute of Theoretical Physics, Chinese Academy of Sciences, Beijing 100190, China}
\affiliation[c]{School of Physical Sciences, University of Chinese Academy of Sciences, No.~19A Yuquan Road, Beijing 100049, China}
\affiliation[d]{Tsung-Dao Lee Institute and School of Physics and Astronomy, Shanghai Jiao Tong University, 800 Dongchuan Road, Shanghai, 200240, China}

\emailAdd{jmli@scu.edu.cn}
\emailAdd{peijunle@mail.itp.ac.cn}
\emailAdd{ranlj@stu.scu.edu.cn}
\emailAdd{zhangwenxing@sjtu.edu.cn}

\abstract{
We study FASER and FASER 2 sensitivities to the quirk signal by simulating the motions of quirks that are travelling through several infrastructures from the ATLAS interaction point to the FASER (2) detector. 
The ionization energy losses for a charged quirk travelling in different materials are treated carefully. 
We calculate the expected numbers of quirk events that can reach the FASER (2) detector for an integrated luminosity of 150 (3000) fb$^{-1}$.
Scenarios for quirks with four different quantum numbers, and different masses and confinement scales are studied. 
}

\maketitle

\section{Introduction}\label{sec:intro} 

After the discovery of the standard model (SM) Higgs boson, the Large Hadron Collider (LHC) is devoted to searching for all kinds of possible new physics beyond the SM. 
However, no unambiguous signs for new physics have been confirmed so far.
There has been a growing interest in exploring scenarios that predict long-lived particles (LLPs)~\cite{Lee:2018pag,Alimena:2019zri,Beacham:2019nyx}. 
The signatures of LLPs are different from those of SM particles, thus were overlooked by traditional searches, {\it e.g. } emerging jet~\cite{Schwaller:2015gea}, disappearing track~\cite{Chen:1996ap,CMS:2018rea}, and kinked track~\cite{Asai:2011wy}. 

Quirks are long-lived exotic particles that are charged under both the Standard Model (SM) gauge group and a new confining gauge group. 
The mass of the lightest quirk is much larger than the confinement scale ($\Lambda$) of the new gauge group, so that two quirks will be connected by a macroscopic gauge flux tube~\cite{Kang:2008ea} when they are produced in pairs. 
Due to the extra long-range gauge interaction, the quirk behaves differently from the SM particles. 
In cases with small confinement scale ($\Lambda < \mathcal{O} (10)$ eV), the new gauge interaction is negligible compared to the magnetic force. 
The hits that a charged quirk leaves inside a detector will be reconstructed as a normal helical track, because finite spacial resolution is considered and the $\chi^2/$DOF as large as 5 is allowed in fitting the track~\cite{CMS:2016ybj}. 
Such signal is found to be constrained by conventional heavy stable charged particle searches at the LHC~\cite{Farina:2017cts}.
When $\Lambda \gtrsim \mathcal{O}(10)$ MeV, the quirk pair system will oscillate intensively after production. Its kinetic energy can be lost quickly via photon and infracolor glueball radiation. Eventually, it will annihilate into the SM particles almost promptly. 
Such quirk signals are found to be constrained by searches for resonances in the SM final states~\cite{Cheung:2008ke,Harnik:2008ax,Harnik:2011mv,Fok:2011yc,Chacko:2015fbc,Capdevilla:2019zbx,Curtin:2021spx}. 
For $\Lambda \in [10~\text{keV},10~\text{MeV}]$, the oscillation amplitude of the quirk is microscopic (un-resolvable by detectors). Meanwhile, the radiation of photon and infracolor glueball is not efficient. 
The electric neutral quirk-pair system will leave hits along a straight line inside the tracker, which will be reconstructed as a single ultra-boosted charged particle with a high ionization energy loss. This signal was searched at the Tevatron~\cite{D0:2010kkd}, giving no evidence. 

The quirk oscillation amplitude becomes macroscopic ($\sim~\text{mm}-\text{m}$) when $\Lambda \sim 100~\text{eV}-\text{keV}$. 
The hits induced by a quirk can no longer be reconstructed as a helical track, will thus be dropped in conventional event reconstruction at the LHC. 
Moreover, due to its heavy mass, the quirk can pass through the calorimeters without losing much energy. So that the missing transverse energy is the signature of quirk and it is constrained by mono-jet searches at the LHC~\cite{Farina:2017cts} when it is recoiling against an energetic initial state radiated (ISR) jet in production.
On the other hand, when the recoiling jet is soft, the quirk-pair system will be stopped in the  calorimeters due to ionization energy loss. Then, after a long time of oscillations the quirk pair annihilates at a time when there are no active $pp$ collisions. Such signal is constrained by the stopped long-lived particles searches at the LHC~\cite{Evans:2018jmd,ATLAS:2013whh,CMS:2017kku}. 
The quirk hits in tracker can also be identified using their coplanarity~\cite{Knapen:2017kly}. Because the infracolor force (for $\Lambda \sim~\text{keV}$) is much larger than the Lorentz force (with magnetic field strength $B\sim \mathcal{O}(1)$ T), the hits of quirk pair distribute approximately on a plane with distance less than $\mathcal{O}(100)$ $\mu$m.  
Moreover, it is found in Ref.~\cite{Li:2019wce} that relatively large ionization energy loss of each quirk hit in the tracker can be used to further improve the coplanar hits search. 

There has been a number of new experiments proposed at CERN, such as FASER~\cite{Feng:2017uoz,FASER:2018ceo,FASER:2018eoc,FASER:2018bac}, MATHUSLA~\cite{Chou:2016lxi,MATHUSLA:2018bqv,Curtin:2018mvb}, and SHiP~\cite{Bonivento:2013jag,SHiP:2015vad,Alekhin:2015byh}. 
They will be able to look for LLPs in different ways from general-purpose detectors such as the ATLAS and the CMS. 
We focus on the FASER (2) experiment in this work, which is located 480 m downstream from the ATLAS interaction point (IP). 
Although FASER is dedicated to searching for light, extremely weakly-interacting particles, such as the dark photon, a charged particle which can pass through all facilities between the ATLAS IP and FASER will also leave visible signal in it. 
The quirk particle produced at the ATLAS IP, because of its large mass and high energy, can reach the FASER (2) detector with a considerable rate. 
The FASER (2) sensitivity to quirk particles is studied, in terms of the quirk quantum numbers, quirk mass and the infracolor confinement scale. 

This paper is organized as follows. In Sec.~\ref{sec2}, we discuss the production of quirks of different quantum numbers at the LHC and give the fraction of events that have quirk pair flying close to the beam axis. Sec.~\ref{sec3} is focused on the equation of motions for quirks, the forward infrastructures from ATLAS interaction point to the FASER detector, and the ionization force on charged particles, which play important roles in our simulation. The results are presented in Sec.~\ref{sec4}, where we talk about the angular momentum of the quirk pair induced by the ionization force, the signal efficiency, the infracolor glueball and electromagnetic radiations, the FASER (2) sensitivity, and features of the quirk signal. We draw the conclusion in Sec.~\ref{sec5}.

\section{Quirk production at the LHC}\label{sec2}

In order to solve the little hierarchy problem~\cite{BasteroGil:2000bw,Bazzocchi:2012de}, some models beyond the Standard Model (BSM) with neutral naturalness~\cite{Curtin:2015bka} predict the existence of color neutral quirk particles. 
Such kind of models includes folded supersymmetry~\cite{Burdman:2006tz,Burdman:2008ek}, quirky little Higgs~\cite{Cai:2008au}, twin Higgs~\cite{Chacko:2005pe,Craig:2015pha,Serra:2019omd,Ahmed:2020hiw}, minimal neutral naturalness model~\cite{Xu:2018ofw} and so on. 
In more general cases, quirk particles can also carry color charge~\cite{Martin:2010kk}, be either fermions or scalar bosons. 

In this work, we will focus on fermionic and scalar quirks that have the same SM quantum numbers as right-handed charged lepton and right-handed down-type quark, {\it i.e.} under $SU(N_{\text{IC}}) \times SU_C(3) \times SU_L(2) \times U_Y(1)$ gauge group, they are 
\begin{align}
	\tilde{\mathcal{D}} &= \left( N_{\text{IC}}, 3, 1, -1/3 \right),~~ \label{eq::qn1}\\
	\tilde{\mathcal{E}} &= \left( N_{\text{IC}}, 1, 1, -1 \right),~~\label{eq::qn2}\\
	\mathcal{D} &= \left( N_{\text{IC}}, 3, 1, -1/3 \right),~~\label{eq::qn3}\\
	\mathcal{E} &= \left( N_{\text{IC}}, 1, 1, -1 \right) \label{eq::qn4},
\end{align}
where $\tilde{\mathcal{D}}^c$ and $\tilde{\mathcal{E}}^c$ are spin zero quirks, $\mathcal{D}^c$ and $\mathcal{E}^c$ are fermionic quirks.
We will take $N_{\text{IC}}=2$ for the infracolor gauge group in this work.  Note that quirk production cross sections are proportional to $N_{\text{IC}}$. 
Although the electric charges of  $\tilde{\mathcal{D}}^c$ and $\mathcal{D}^c$ are $-\frac{1}{3}$, one can only observe the quirk-quark bound state with integral electric charges due to the color confinement. 
The probability for the quirk-quark bound state pair to have charge $\pm 1$ is around 30\%~\cite{Knapen:2017kly}.
In the following discussions, we will simply refer to the charge $\pm 1$ quirk-quark bound state as  $\tilde{\mathcal{D}}^c$ or $\mathcal{D}^c$ because only final states with non-zero electric charges are concerned. 
Moreover, the equation of motions (EoM) for the quirk-quark bound states are similar as the EoM for the quirks since the masses of the quirks are much larger than those of the quarks. 

\begin{figure}[thb]
	\begin{center}
		\includegraphics[width=0.32\textwidth]{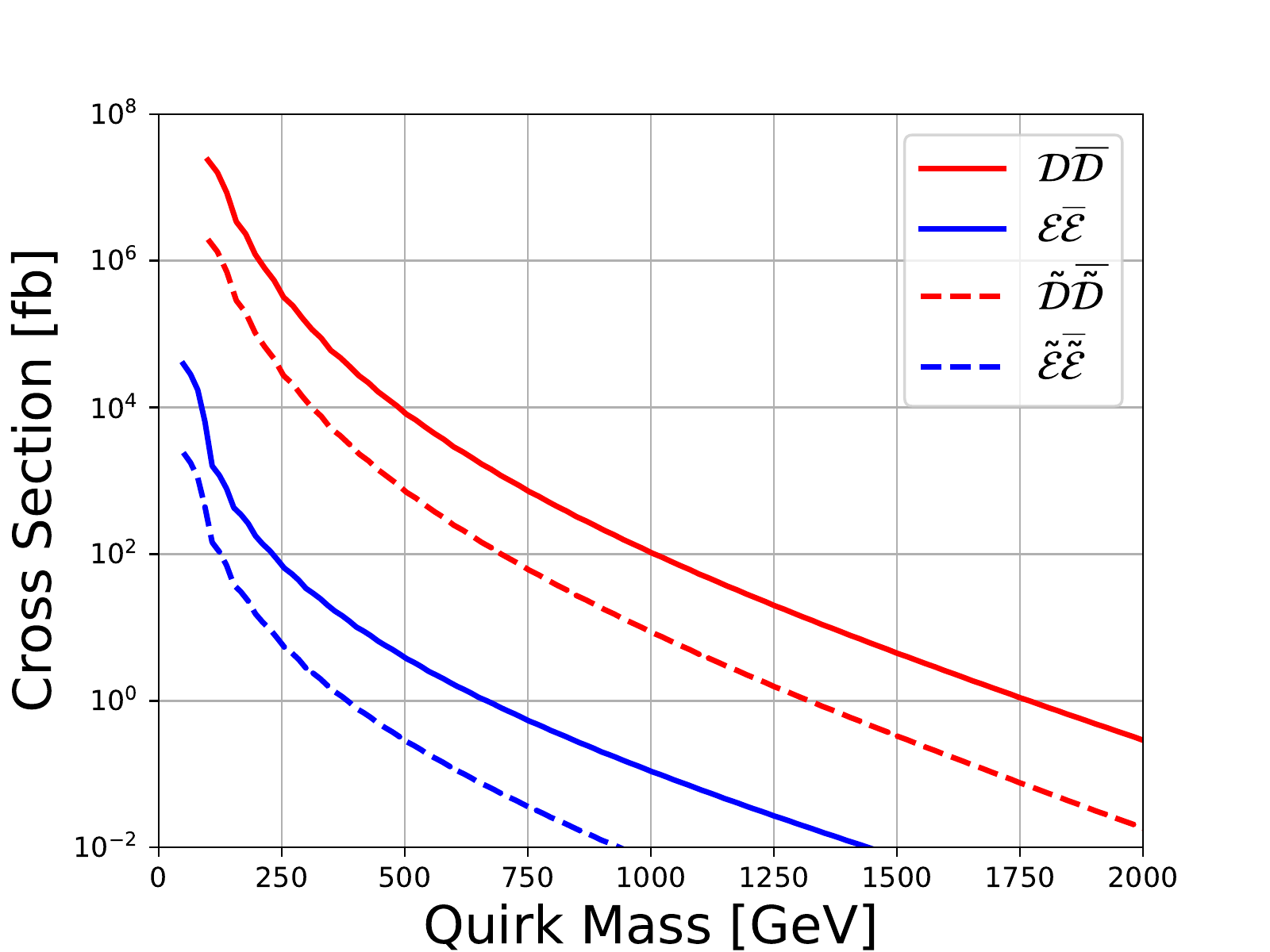}
		\includegraphics[width=0.32\textwidth]{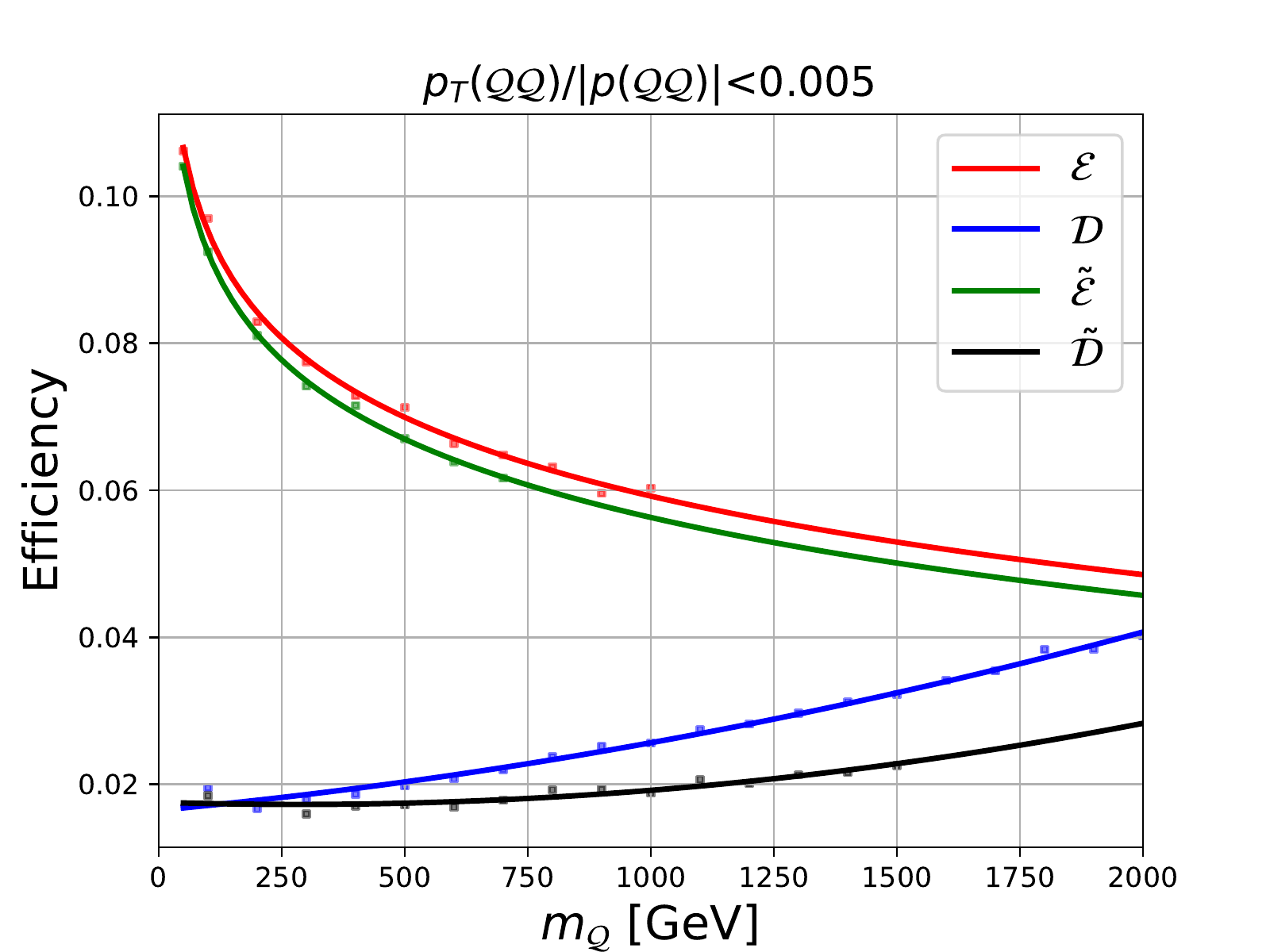}
		\includegraphics[width=0.32\textwidth]{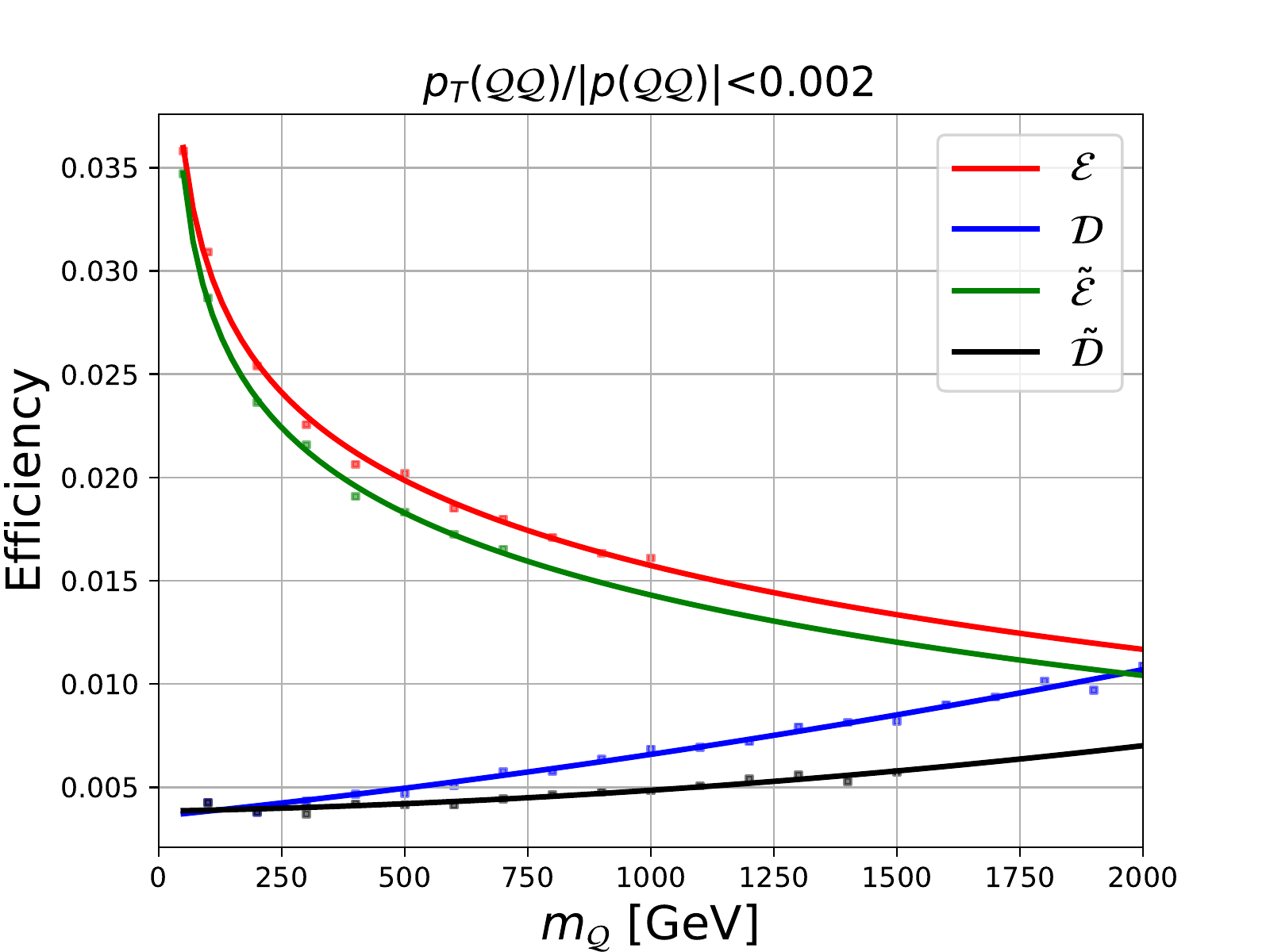}
	\end{center}
	\caption{The leading order production cross sections for the quirk pair production at the 13 TeV LHC (left panel). The fraction of events that have $p_T(\mathcal{Q}\mathcal{Q})/ |p(\mathcal{Q}\mathcal{Q})|< 0.005$ (middle panel) and $0.002$ (right panel) for different quirks and their masses. }
	\label{fig::xsecs}
\end{figure}

The quirks can be pair produced at colliders through the SM gauge interactions. 
The colored quirks productions are dominated by the gluon initialed processes while the color neutral quirks productions are given by the Drell-Yan processes. 
The leading order production cross section (calculated by MG5\_aMC@NLO~\cite{Alwall:2014hca}) for each quirk is shown in the left panel of Figure~\ref{fig::xsecs}. 
Because of less degree of freedom and momentum suppressed couplings to the gauge bosons, given the same mass and quantum numbers, the scalar quirk has much smaller production cross section than the fermionic quirk.   

Although each quirk can be produced with large transverse momentum, the quirk-pair system ($\mathcal{Q}\mathcal{Q}$) is travelling along the beam axis (thus enters the FASER (2) detector) if the initial state and final state radiations are not taken into account. 
In event generation, the effects of initial state radiation (ISR), final state radiation (FSR), and the hadronization of colored final state are simulated with Pythia8~\cite{Sjostrand:2007gs}. 
Those effects deflect the direction of the $\mathcal{Q}\mathcal{Q}$ system from the beam axis, making many of the quirk production events undetectable by the FASER detector~\footnote{In this work, all quirk events are simulated at the leading order (LO) with additional jets generated by parton shower. The next-to-LO QCD corrections can increase the event production rates in the lower $p_T(\mathcal{Q}\mathcal{Q})$ region for both colored and color neutral quirks as have been studied in Refs.~\cite{Cullen:2012eh,Backovic:2015soa,Ruiz:2015zca,Fuks:2016ftf} for similar scenarios, leading to improved FASER (2) sensitivities.}.  
In the middle and right panels of Figure~\ref{fig::xsecs}, we plot the fractions of events that have $p_T(\mathcal{Q}\mathcal{Q})/ |p(\mathcal{Q}\mathcal{Q})|< 0.005$ and 0.002 for different quirks, where $p_T(\mathcal{Q}\mathcal{Q})$ and $|p(\mathcal{Q}\mathcal{Q})|$ are the transverse momentum and the momentum size of the quirk-pair system, respectively. 
Note that these selections also keep events with $\mathcal{Q}\mathcal{Q}$ system travelling opposite to the $Z$-axis, which obviously can not reach FASER (2). 
From the figure, we can observe that there are $\sim 2-10\%$ events of $\mathcal{Q}\mathcal{Q}$ flying around the beam axis ($p_T(\mathcal{Q}\mathcal{Q})/ |p(\mathcal{Q}\mathcal{Q})|< 0.005$) after including the ISR and FSR effects. And the fraction is reduced by a factor of 4 for more stringent condition ($p_T(\mathcal{Q}\mathcal{Q})/ |p(\mathcal{Q}\mathcal{Q})|< 0.002$). 
The ISR and FSR are much more intensive for the colored quirk production processes than the color neutral ones, thus the fraction of the events that pass the deflection condition is lower for colored quirks. 
For colored quirk production, the FSR dominates over the ISR. Heavier quirk is more difficult to be deflected by the FSR, so the selection efficiency is higher for heavier quirk. 
While for the color neutral quirk, only the ISR is important. And the energy scale of the ISR is proportional to the quirk mass, which means harder ISR will occur for heavier quirk, leading to lower selection efficiency. 
The difference between the fermionic quirk and scalar quirk is owning to the fact that the phase space with larger momentum transfer is enhanced for scalar quirk production due to its momentum dependent coupling with the SM gauge bosons.

\section{Traveling toward  FASER (2)}\label{sec3}

When the quirk pair moves toward the FASER detector through materials, the motion of each quirk is controlled by~\cite{Kang:2008ea}
\begin{align}
	\frac{\partial ({m} \gamma \vec{v})}{\partial t}
	&=\vec{F}_{s}+\vec{F}_{\text{ion}}~,\label{eq::move}\\
	\vec{F}_{s}&=-\Lambda^2\sqrt{1-\vec{v}_{\perp}^{2}} \hat{s}-\Lambda^2 \frac{v_{ \|} \vec{v}_{\perp}}{\sqrt{1-\vec{v}_{\perp}^{2}}}~,\label{eq::fs}\\
	\vec{F}_{\text{ion}}&= \frac{dE}{dx}\hat{v}~, \label{fmc}
\end{align}
where {$\gamma=1/\sqrt{1-\vec{v}^2}$,} $v_{ \|}=\vec{v}\cdot\hat{s}\nonumber$ and $\vec{v}_{\perp}=\vec{v}-v_{ \|}\hat{s}\nonumber$ with $\hat{s}$ being a unit vector along the string pointing outward at the endpoints. $\vec{F}_s$ corresponds to the infracolor force and is described by the Nambu-Goto action, and $\Lambda$ is the confinement scale. $\vec{F}_{\text{ion}}$ represents the force arising from the effects of ionization energy loss for charged quirk propagating through materials. Note that we have ignored several sub-dominating energy loss effects such as infracolor glueball and photon radiations. 
Moreover, a dedicated simulation of R-hadron propagation inside detector by experimentalists~\cite{ATLAS:2013whh} shows that the energy lost through hadronic interactions is much smaller than that through electromagnetic ionization. So, we do not consider this effect in the quirk EoM.

To solve Eq.~\ref{eq::move}, we have to consider both the quirk-pair centre of mass (CoM) frame and the laboratory (lab) frame. In the CoM frame, $\hat{s}$ is approximately parallel to the vector difference between positions of the two quirks (this is only true for $\Lambda^2 \gg F_{\text{ion}}$, see Ref.~\cite{Kang:2008ea}). However, the CoM frame itself is changing all the time due to effects of $\vec{F}_{\text{ion}}$, which is related to the quirk velocity in the lab frame. The procedures of numerically solving the EoM by slowly increasing the time with small steps were introduced in Ref.~\cite{Li:2019wce}.

The mean rate of energy loss of moderately relativistic ($0.1 \lesssim \beta \gamma \lesssim 1000$) charged heavy particles is well described by the Bethe-Bloch (BB) formula \cite{ParticleDataGroup:2020ssz},
\begin{align}\label{mcbb}
	\left\langle-\frac{d E}{d x}\right\rangle_{\text{BB}}=K\rho z^{2} \frac{Z}{A} \frac{1}{\beta^{2}}\left[\frac{1}{2} \ln \frac{2 m_{e} c^{2} \beta^{2} \gamma^{2} W_{\max }}{I^{2}}-\beta^{2}-\frac{\delta(\beta \gamma)}{2}\right]~, 
\end{align}
where $\rho$, $Z$, $A$, and $I$ are the density, the atomic number, the relative atomic mass, and the mean excitation energy of the material, respectively. $z$ and $\beta$ are the charge number and the velocity of the incident particle, and $\gamma=\sqrt{1-\beta^2}$. $m_e$ is the electron mass and $K=0.307$ MeV $\text{mol}^{-1}$ $\text{cm}^2$. $W_{\text{max}}$ stands for the maximum possible energy transfer to an electron in a single collision, which is given by \cite{ParticleDataGroup:2020ssz}
\begin{align}
	W_{\max }=\frac{2 m_{e} c^{2} \beta^{2} \gamma^{2}}{1+2 \gamma m_{e} / M+\left(m_{e} / M\right)^{2}}~,
\end{align}
where $M$ is the mass of the incident particle. The density effect correction to the ionization energy loss is cast into $\delta(\beta\gamma)$, which is usually computed using Sternheimer's parameterization as \cite{ParticleDataGroup:2020ssz}
\begin{align}
	\delta(\beta \gamma)=\left\{\begin{array}{ll}
		2(\ln 10) x-\bar{C} & \text { if } x \geq x_{1} \\
		2(\ln 10) x-\bar{C}+a\left(x_{1}-x\right)^{k} & \text { if } x_{0} \leq x<x_{1} \\
		0 & \text { if } x<x_{0} \text { (nonconductors) } \\
		\delta_{0} 10^{2\left(x-x_{0}\right)} & \text { if } x<x_{0} \text { (conductors) }
	\end{array}\right.~,
\end{align}
where $x=\log_{10}(\beta\gamma)$, $\bar{C}=1-2\ln\left(\frac{\hbar\omega_p}{I} \right)$, and $\hbar\omega_p=\sqrt{\rho \langle Z/A\rangle}$. 

\begin{figure}[thb]
	\begin{center}
		\includegraphics[width=1.0\textwidth]{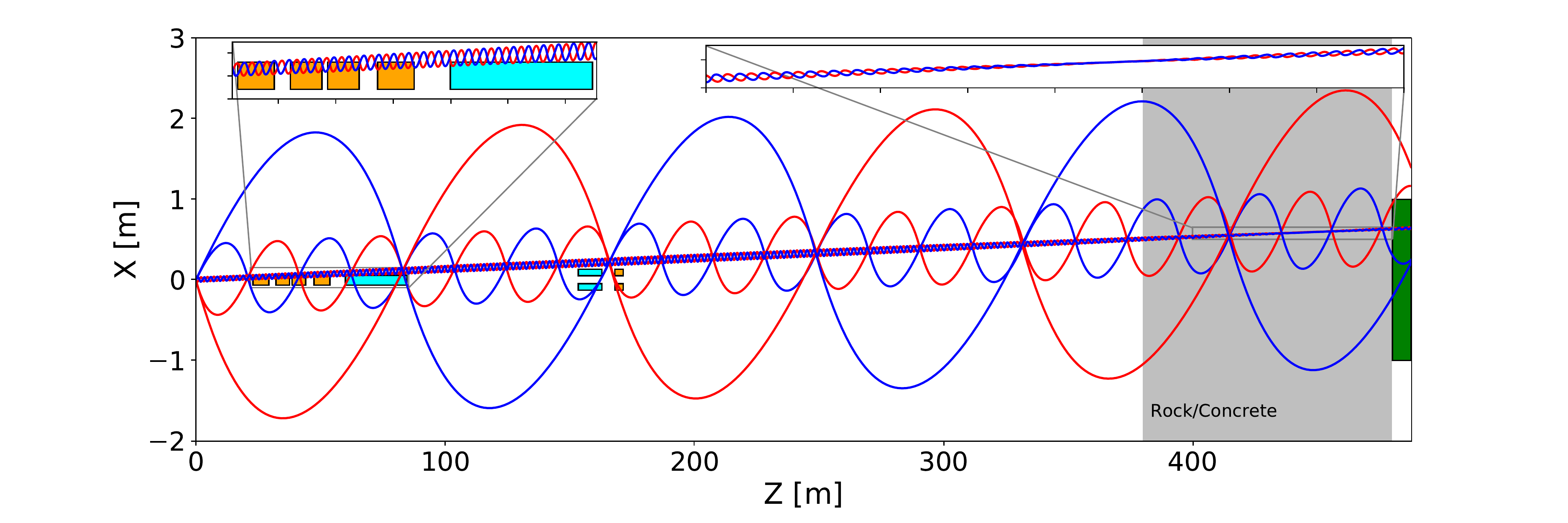}
	\end{center}
	\caption{The quirk trajectories and forward infrastructures from ATLAS IP to the FASER (2) detector. Two fragments of quirk trajectory are magnified for clearer visibility. The quirk initial momenta are $\vec{p}_1 =(-132.146, 121.085, 1167.35)$ GeV and $\vec{p}_2 =(136.381, -123.865, 2061.56)$ GeV and the quirk mass is 800 GeV. And three different confinement scales $\Lambda=50~\text{eV}, 100~\text{eV}, 400~\text{eV}$ are considered for illustration. The orange, cyan, grey, and green regions indicate the regions with quadrupole magnetic field, dipole magnetic field, rock/concrete and the FASER 2 detector, respectively. }
	\label{fig::quirktrajectory}
\end{figure}

\begin{figure}[thb]
	\begin{center}
		\includegraphics[width=0.45\textwidth]{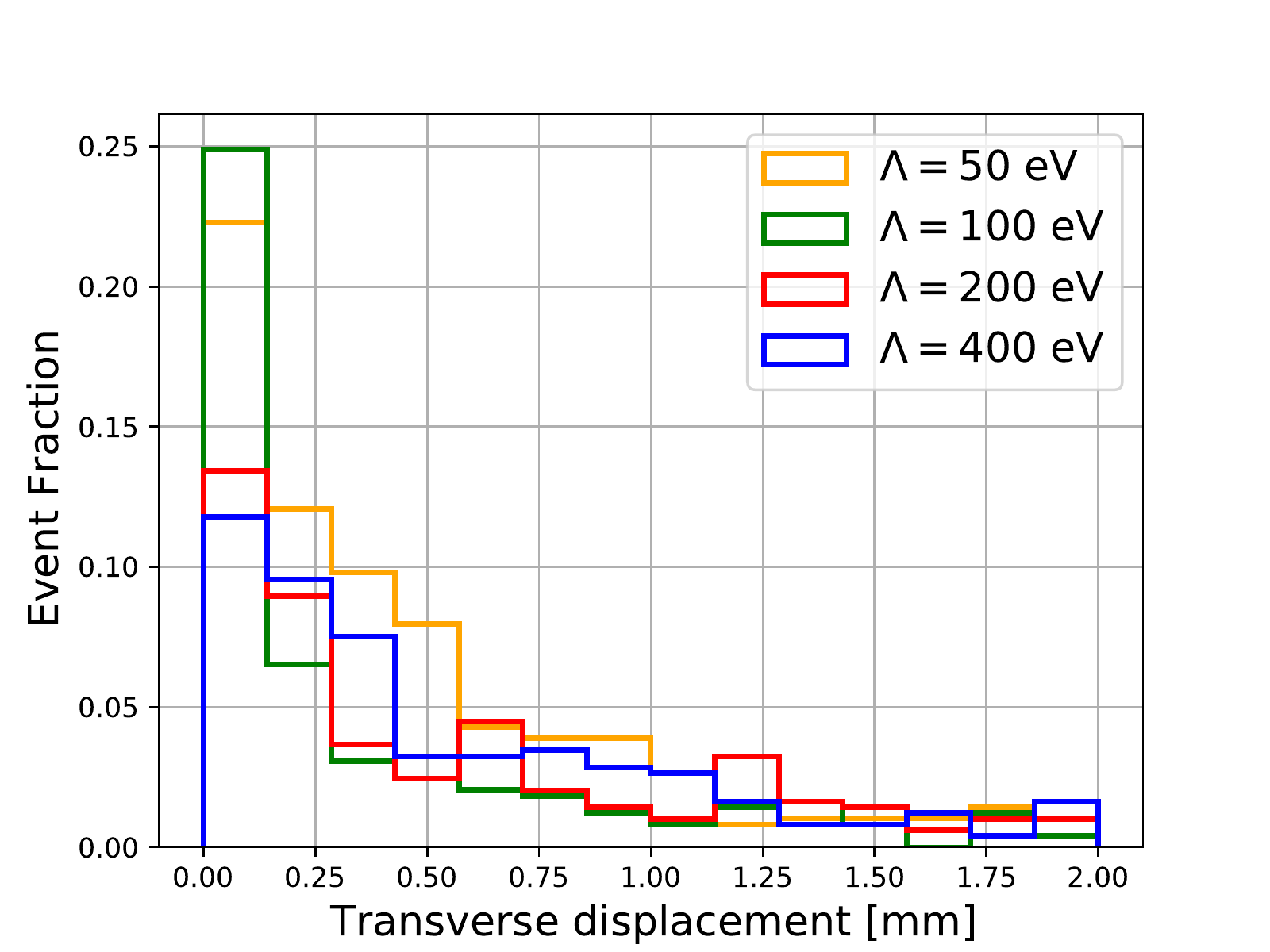}
		\includegraphics[width=0.45\textwidth]{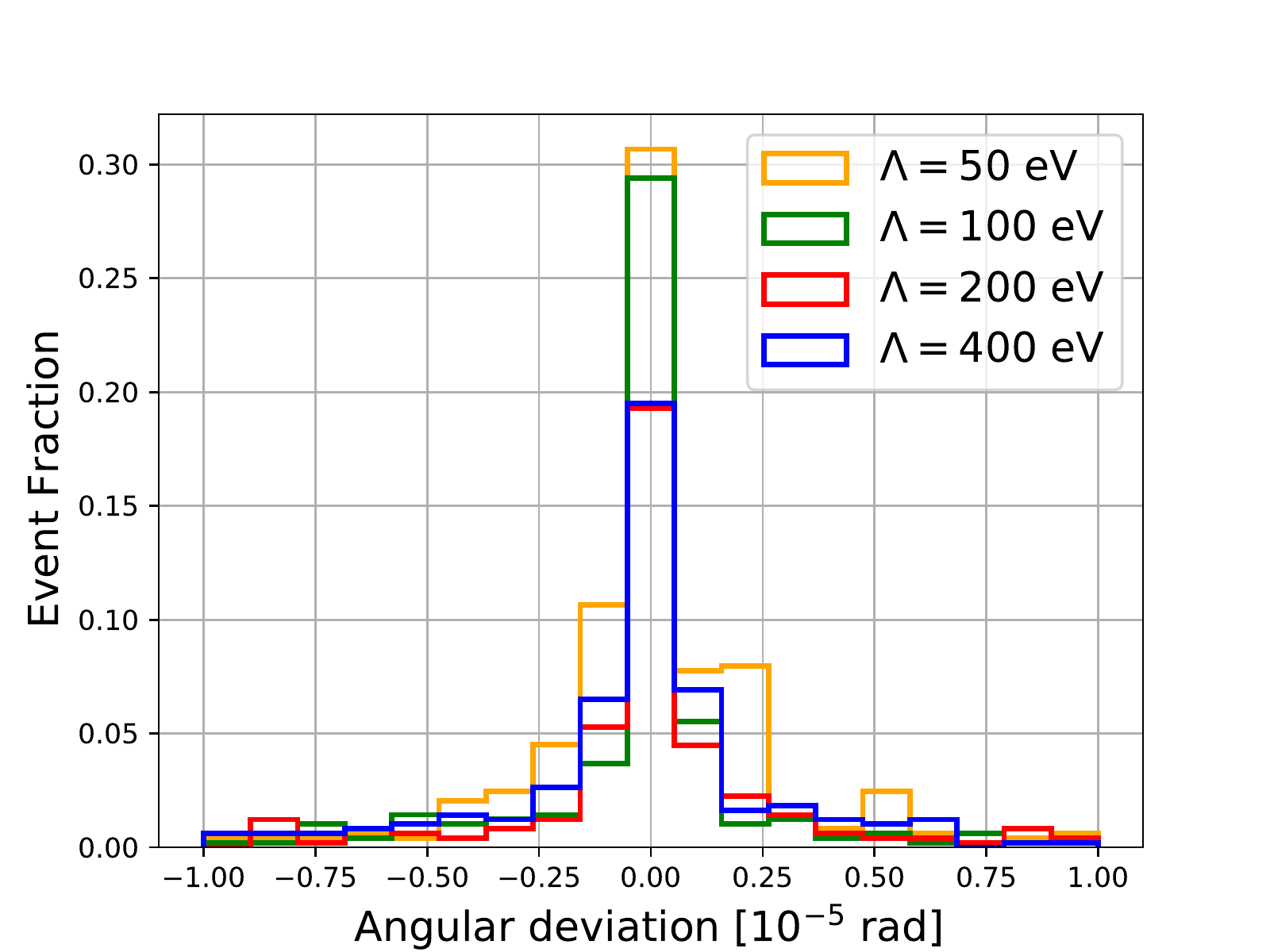}
	\end{center}
	\caption{The distributions of transverse displacement and angular deviation of the quirk-pair system at the first crossing point after the D2 magnets, due to the effects of D1 and D2. Events of the $\mathcal{E}$ quirk with mass 300 GeV are used. }
	\label{fig::mag}
\end{figure}

A schematic drawing of quirk trajectories from the ATLAS IP to the FASER 2 detector is showing in Figure~\ref{fig::quirktrajectory}. The regions with quadrupole magnetic field and dipole magnetic field are indicated by orange and cyan rectangles, respectively. 
	The quadrupole magnetic field at the LHC is produced by several component magnets~\cite{Evans:2008zzb,Adamczyk:2015cjy}. 
	Each quadrupole magnet can only focus charged particles in one direction and defocus them in the orthogonal direction. However, the alternated focusing and defocusing quadrupole magnets in the magnet lattice lead to a net effect of focusing. For example, the inner triplet (indicated by the orange rectangle at $z \sim [20,50]$ m) on the right side of the ATLAS detector is made up of four quadrupole magnets. Among them, Q1 and Q3 are focusing while Q2a and Q2b are defocusing~\cite{USLHC:2006ebl,Devred:2000te}. 
	As a result, the quadrupole magnetic fields are likely to push the quirks toward the forward direction, giving higher signal rates at FASER (2). 
	In Figure~\ref{fig::mag}, we show the distributions of transverse displacement and angular deviation of the quirk-pair system at the first crossing point (which is at around $z=200$ m for $\Lambda \in[50,400]$ eV) after the D2 magnets, due to the effects of D1 and D2 (the specifications of them are given in Table~\ref{material}). 
	Events of the $\mathcal{E}$ quirk with mass 300 GeV have been used for illustration. The distributions are mainly controlled by $\Lambda$ and are similar for other quirks with different masses. In the parameter region of our interest, the transverse displacements are less than 1 mm and the angular deviations are smaller than $2 \times 10^{-5}$ rad for most of the events. 
	In the following, the effects of magnetic fields are ignored for simplicity.

We will focus on the ionization effects of quirk travelling down-stream from the ATLAS IP, which not only leads to energy loss of each quirk, but also induces the angular momentum for the quirk-pair system.  
The configurations of infrastructures between the ATLAS IP and the FASER detector used in the simulation of this work are summarized in Table~\ref{material}~\cite{Adriani:926196}. 
The quadrupole magnetic fields are ignored according to the discussions given above. 
Moreover, we do not consider the sizes of magnets and other electric devices in the tunnel, since their effects should be small compared to the 100-meter rock and concrete. 
{The TAS is a 1.8-meter-long copper block with an inner radius of 17 mm, which is put at 19 m downstream. Note that we assume an infinite transverse size of the TAS, which is also used to represent other materials at around the beginning of quirks' motion. 
	The TAN at 140 m downstream contains a 9.6 cm wide by 100 cm long by 60.7 cm deep slot, which is occupied by copper bars.  
	Finally, the quirk needs to travel through 10 m of concrete and 90 m of rock before reaching FASER (2) at 480 m downstream. 
	FASER contains three tracking stations and each tracking station is made up of three tracking planes of size $0.32\times 0.32~\text{m}^2$. In the simulation of this work, FASER 2 also contains nine tracking planes in three tracking stations and each tracking plane for FASER 2 is chosen to have a size of $2\times 2~\text{m}^2$.
}

\begin{table}[htb]
	\centering
	\begin{tabular}{ccc}  
	\hline
	Component  & $x,y,R$[m] & $z$[m] \\
	\hline 
	TAS (Copper) & $R>0.017$ & 19$-$20.8 \\
	D1 (3.5 T) & $R<0.06$  &  59.92$-$84.65 \\
	TAN (Copper) &  $|x|<0.047$,  $-0.538<y<0.067$ &  140$-$141 \\
	D2 (3.5 T) & $(x\pm 0.093)^2+y^2<0.04^2$  &  153.48$-$162.93 \\
	Concrete & $R>0$  &  380$-$390 \\
	Rock &  $R>0$ &  390$-$480\\
	Tracker of FASER & $|x|<0.16$,  $|y|<0.16$  &  $|z-481.6/482.8/484.0|<0.041$ \\
	Tracker of FASER 2 & $|x|<1$,  $|y|<1$  &  $|z-485.1/486.3/487.5|<0.041$ \\
	\hline
\end{tabular}
	\caption{\label{material} The configurations of infrastructures between the ATLAS IP and the FASER (2) detector. The ATLAS IP is the ordinate origin and the transverse distance is $R=\sqrt{x^2+y^2}$.}
\end{table}

According to Ref.~\cite{Groom:2001kq}, values of the variables relevant to the ionization energy loss in copper, concrete, and rock are listed in the Table \ref{varbs}. It is noted that the values of $a$, $k$, $x_0$, $x_1$, $\bar{C}$, and $\delta_0$ are obtained from the muon travelling through materials. We use them to estimate the ionization energy loss of the quirk in the corresponding materials.

\begin{table}[htb]
	\centering
	\scriptsize{\begin{tabular}{ccccccccccc}  
		\hline
		Material & $Z$ or $\langle Z\rangle$ & $\langle Z/A\rangle$[mol/g] & $\rho$[g/$\text{cm}^3$] & $I$[eV]  &$a$ & $k$ & $x_0$ & $x_1$ & $\bar{C}$ & $\delta_0$\\
		\hline 
		Copper &29 &29/63.546 & 8.960 & 322.0 & 0.14339 & 2.9044 & -0.0254 & 3.2792 & 4.4190 & 0.08\\
		Concrete & 8.56 &0.50274 & 2.300    & 135.2 & 0.07515 & 3.5467 & 0.1301  & 3.0466 & 3.9464 & 0.00\\
		Rock & 11 & 0.50000 & 2.650     & 136.4 & 0.08301 & 3.4120 & 0.0492  & 3.0549 & 3.7738 & 0.00\\
		\hline
	\end{tabular}}
	\caption{\label{varbs} Values of the variables relevant to the ionization energy loss in copper, concrete, and rock. The values of $a$, $k$, $x_0$, $x_1$, $\bar{C}$, and $\delta_0$ are for the muons~\cite{Groom:2001kq}.}
\end{table}

\begin{figure}[thb]
	\begin{center}
		\includegraphics[height=0.5\textwidth]{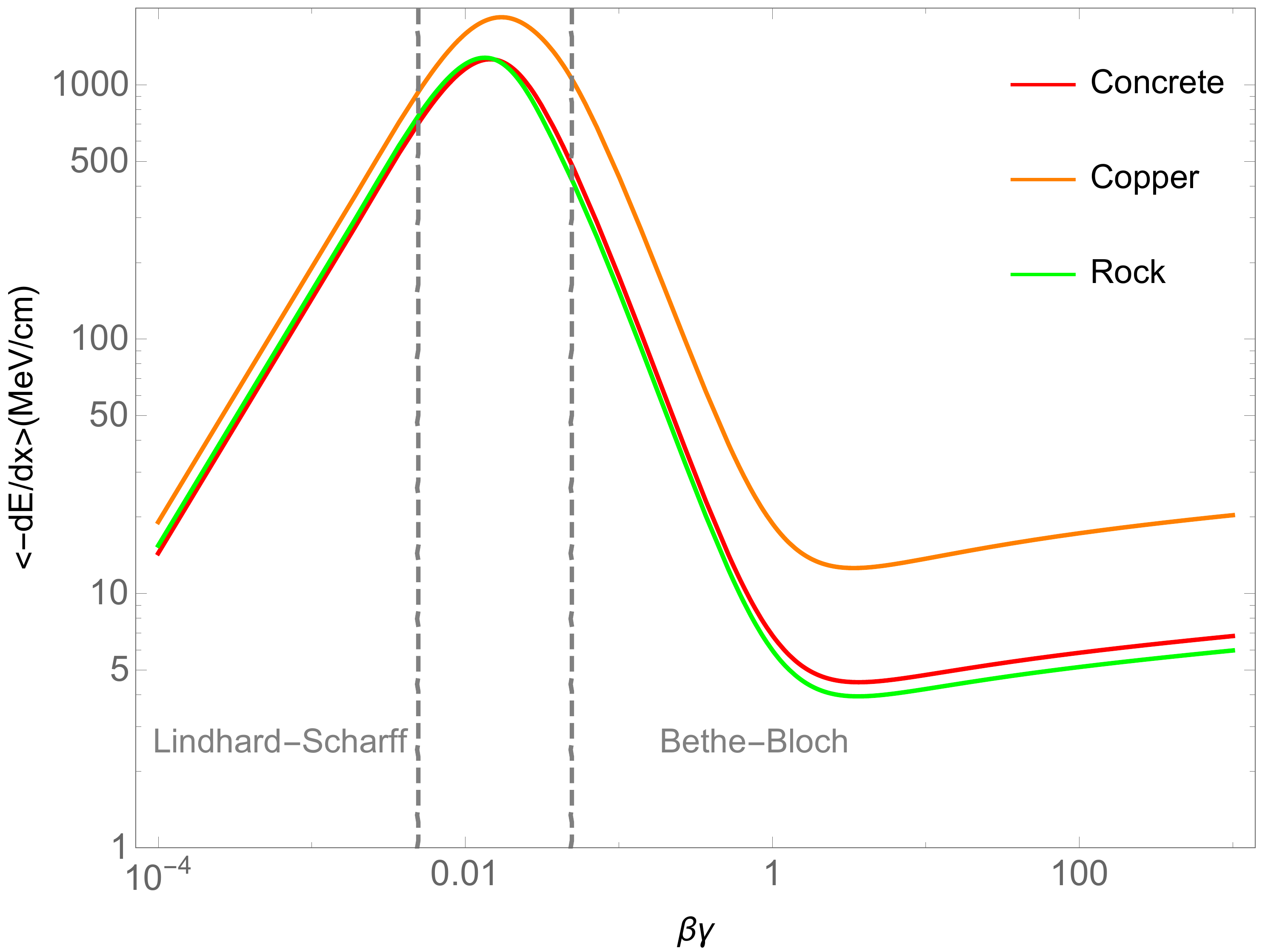}
	\end{center}
	\caption{\label{deodx} The mean rates of energy loss for charged particle traveling through concrete, copper, and rock, supposing $z=1$ and $m_e/M\ll1$.}
\end{figure}

In the region $v/c<\left( 7.33\times 10^{-3}\right) z^{2/3}$, the mean rate of energy loss is described by the Lindhard-Scharff (LS) formula,
\begin{align}\label{mcls}
	\left\langle-\frac{d E}{d x}\right\rangle_{\text{LS}}=3.1\times 10^{-11} \text{GeV}^2
	\frac{\rho}{\text{g}/\text{cm}^3}\frac{z^{7/6}Z/A}{\left(z^{2/3}+Z^{2/3} \right)^{3/2} }\beta~.   
\end{align}
In Figure~\ref{deodx}, the mean rates of energy loss  for charged particle travelling through concrete, copper, and rock are plotted, where we set $z=1$. The $\langle -dE/dx \rangle$ is independent of the mass of the incident particle because $m_e/M\ll 1$ is assumed. It is noted that $\langle -dE/dx \rangle$ values between LS and BB regions are obtained by interpolation. 

When solving the quirk EoM numerically, the mean energy loss $\left\langle-\frac{d E}{d x}\right\rangle$ can not be used directly for estimating $| \vec{F}_{\text{ion}}|$. 
The ionization energy loss for a charged particle travelling a distance $\delta_x$ in the material fluctuates, which, in the BB region, can be described by a Gaussian distribution when $\delta_x$ is large enough such that $\frac{\xi(\delta_x)}{W_{\text{max}}}>10$~\cite{ParticleDataGroup:2020ssz}, where
\begin{align}
	\xi(\delta_x)=\frac{K\rho z^{2}Z}{2A\beta^{2}}\delta_x~.
\end{align} 
This corresponds to our case, because the quirks travel through macroscopic region of materials with $v/c \sim 1$. 
The $-dE/dx$ for each quirk satisfies the Gaussian distribution with probability distribution function (PDF)~\cite{ParticleDataGroup:2020ssz}
\begin{align}\label{pdf}
	f\left( -\frac{dE}{dx}, \delta_x\right) &= \frac{1}{\sqrt{2\pi }\sigma(\delta_x)}\exp\left[-\frac{\left(-\frac{dE}{dx}-\left\langle-\frac{d E}{d x}\right\rangle_{\text{BB}}\right)^2 }{2\sigma^2(\delta_x)} \right]~,\\
	\sigma^2(\delta_x) &=\frac{\xi(\delta_x)W_{\text{max}}}{\delta_x^2}(1-\beta^2/2)~.
\end{align}
At each time grid of our simulation, the $-dE/dx$ is randomly generated by the PDF in Eq. \ref{pdf} and the $\vec{F}_{\text{ion}}$ is calculated by Eq.~\ref{fmc}.

\section{Results}\label{sec4}

\subsection{Angular momentum induced by the ionization force}

{
	Without the $\vec{F}_{\text{ion}}$, the quirk-pair system travels along a straight line and the trajectories of two quirks lie in the same plane. 
	The ionization force on quirks will change their trajectories, which can be observed in Figure~\ref{fig::quirktrajectory} where the trajectories of the quirk pair between the ATLAS IP and FASER (2) has been twisted when $\Lambda=400$ eV. The twisted trajectories indicate that the direction of the oscillation of the quirk pair keeps changing in the CoM frame due to the non-zero angular momentum. 
	
	As pointed out above, the effects of ionization energy loss for charged quirk propagating through materials will make the trajectory of the quirk deviate from the one obtained by neglecting $\vec{F}_{\text{ion}}$. What is more, the angular momentum of the quirk-pair system will be changing under the influence of $\vec{F}_\text{ion}$ induced by the materials between the IP and the FASER (2) detector. The angular momentum of the quirk-pair system in the lab frame is defined as 
	\begin{align}
		\vec{L}_{tot}=\vec{r}_1\times\vec{p}_1+\vec{r}_2\times\vec{p}_2~.
	\end{align}
	After boosting to the CoM frame, we denote the angular momentum of the quirk-pair system as $\vec{L}_{tot}^\prime$. The initial $\vec{L}_{tot}^\prime$ at the IP is zero. If the quirk pair travels without the impact of $\vec{F}_\text{ion}$ after production, the $\vec{L}_{tot}^\prime$ will keep being as zero, which means that the two quirks always oscillate along the same line in the CoM frame. At some moment the quirks inject into the material and the impact of $\vec{F}_\text{ion}$ is turned on, then the $\vec{L}_{tot}^\prime$ obtains a vaule. Without losing generality, we demonstrate how the angular momentum of the quirk-pair system changes by analysing the forces on one quirk from the pair. We denote the velocity of the quirk pair at this moment as $\vec{\beta}$ ($|\vec{\beta}|=\beta$). The velocity of the quirk is $\vec{v}$ ($|\vec{v}|=v$). The components of $\vec{v}$ parallel and perpendicular to $\vec{\beta}$ are denoted as $v_p$ and $v_c$, respectively. $\vec{F}$ ($|\vec{F}|=F$) is the force on the quirk due to the the effects of ionization energy loss, whose components parallel and perpendicular to $\vec{\beta}$ are denoted as $F_p$ and $F_c$, respectively. Because $\vec{F}$ is in the same direction as $-\vec{v}$, we have 
	\begin{align}
		&F_p=\frac{v_p}{v}F~,\\
		&F_c=\frac{v_c}{v}F~.
	\end{align}   
	After boosting to the CoM frame, it is not hard to get
	\begin{align}
		&\frac{F_p^\prime}{v_p^\prime}=\frac{F}{v}\frac{v_p-\beta v}{v_p-\beta}~,\\
		&\frac{F_c^\prime}{v_c^\prime}=\frac{F}{v}~,
	\end{align}   
	where the primed variables are in the CoM frame. 
	Since $\frac{F_p^\prime}{v_p^\prime}\neq \frac{F_c^\prime}{v_c^\prime}$, $\vec{F}^\prime$ has non-zero component perpendicular to $\vec{v}^\prime$, which induces no-zero $\vec{L}_{tot}^\prime$. 
	
	When $\Lambda^2\gg F$ is satisfied, the position and velocity of each quirk from the pair in the CoM frame are approximately given by \cite{Li:2020aoq}
	\begin{align}
		r^\prime(g) &=\frac{m}{\Lambda^{2}}\left(\sqrt{1+\rho^{2}}-\sqrt{1+\rho^{2}(1-2 g)^{2}}\right)~,\label{rcom} \\ 
		v^\prime(g) &=\frac{\rho(1-2 g)}{\sqrt{1+\rho^{2}(1-2 g)^{2}}}~, \label{vcom}
	\end{align}  	 
	respectively, where $\rho=|\vec{P}^\prime|/m$ and $g=\frac{\Lambda^{2} t^\prime}{2 m \rho}$ with $\vec{P}^\prime$ and $t^\prime$ standing for the initial quirk momentum and the time in the CoM frame, respectively. The $\rho$ varying from 0 to 1 corresponds to a period between the two quirks leaving each other and their next meet. In the CoM frame, we denote the angle between $\vec{\beta}$ and the direction of the oscillation of the quirk pair as $\theta$. In the lab frame we get
	\begin{align}
		&v_{p1/2}=\frac{\beta\pm v^\prime(g)\cos\theta}{1\pm\beta v^\prime(g)\cos\theta}~,\\
		&v_{c1/2}=\pm v^\prime(g)\sin\theta \frac{1-\beta v_{p1/2}}{\sqrt{1-\beta^2}}~,\\
		&v_{i}=\sqrt{v_{pi}^2+v_{ci}^2}~~~~~i=1,2.
	\end{align}  
	So, the torque on the $i$-th quirk in the CoM frame is
	\begin{align}
		&T^\prime_i=\frac{v^\prime(g)\cos\theta}{v_i}\left( \frac{v_{pi}-\beta v_i}{v_{pi}-\beta}-1\right) r^\prime(g)F_i(v_i)\sin\theta~~~~~i=1,2, \label{ti}
	\end{align} 
	and the total torque on the quirk-pair system is
	\begin{align}
		&T^\prime_{tot}=T^\prime_1+T^\prime_2~. \label{ttot}
	\end{align} 
	
	The left panel of Figure~\ref{fig::torq} shows the torques on a pair of quirks in the CoM frame based on Eq.~\ref{ti} and Eq.~\ref{ttot} when the quirk pair is travelling through the material of copper during a complete oscillation ($\rho$ ranging from 0 to 1). It is noted that we have used the mean rate of energy loss from Eq.~\ref{mcbb} and Eq.~\ref{mcls} for $F_{i}(v_i)$ ($i=1,2$) without considering the Gaussian PDF in the left panel of Figure~\ref{fig::torq}. We find that $\vec{L}^\prime_{tot}$ keeps almost unchanged after a complete oscillation in the material due to
	\begin{align}
		\int_{0}^{1} (T_1^\prime+T_2^\prime)dg\approx 0~.
	\end{align}
	So, it is concluded that $\vec{L}^\prime_{tot}$ changes sharply only at the moment that the quirk pair injects into or moves out of the material, or just one quirk from the pair is travelling in the material.
	
	\begin{figure}[thb]
		\begin{center}
			\includegraphics[width=0.4\textwidth]{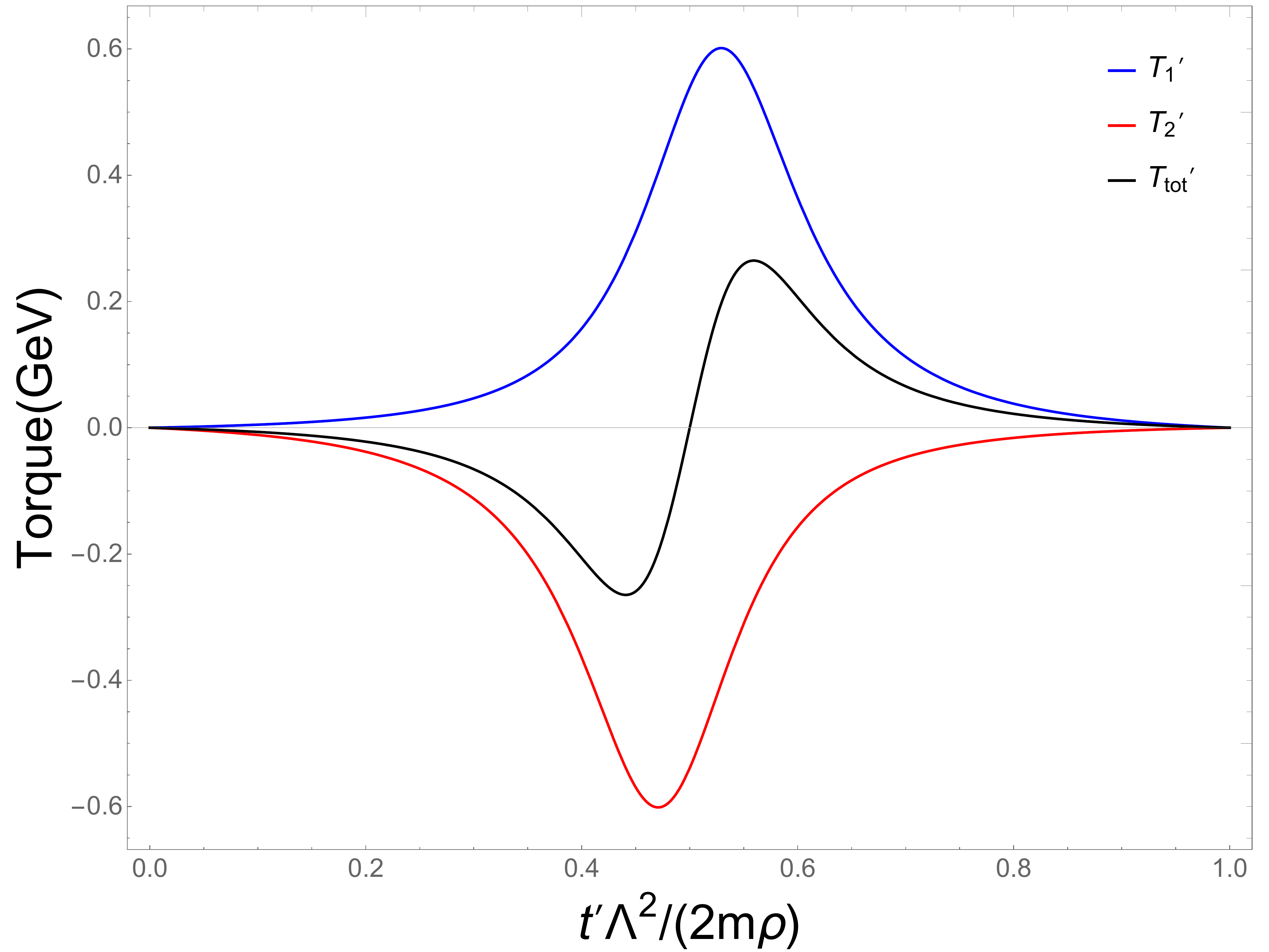}
			\includegraphics[width=0.45\textwidth]{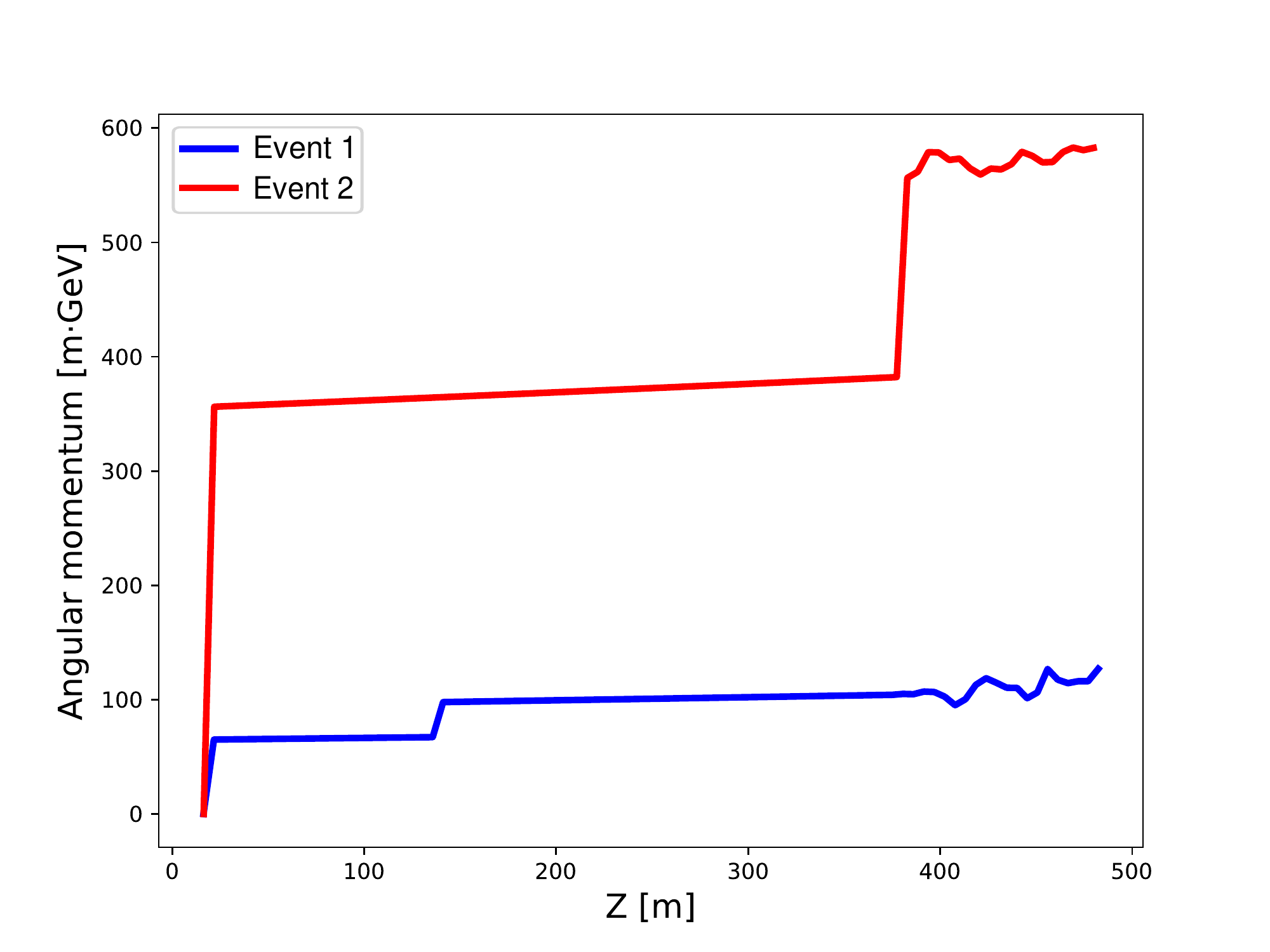}
		\end{center}
		\caption{Left: Torques on quirks in the CoM frame when the quirk pair is travelling in the material of copper. The Gaussian PDF of the energy loss is ignored and the relevant parameters are set as $m=100$ GeV, $z=1$, $\Lambda=300$ eV, $\rho=5$, $\beta=0.9$, and $\theta=\pi/3$.
			Right: The angular momenta of two quirk pairs in the CoM frame when travelling from the IP to the FASER (2) detector. The Gaussian PDF of the energy loss is considered. We have set $\Lambda=300$ eV.  Blue line corresponds to initial momenta $\vec{p}_1=(297.291, -620.587, 1840.02)$ GeV, $\vec{p}_2=(-295.778, 617.359, 1959.69)$ GeV, and red line corresponds to initial momenta $\vec{p}_1=(12.2825, -480.523, 2910.31)$ GeV, $\vec{p}_2=(-10.7928, 485.805, 1086.32)$ GeV.}
		\label{fig::torq}
	\end{figure}

	The right panel of Figure~\ref{fig::torq} shows the angular momenta of two quirk-pair systems in the CoM frame ($\vec{L}_{tot}^\prime$) with respect to $Z$-axis. For both events, the first leap of $\vec{L}_{tot}^\prime$ takes place at $z\sim$ 19 m when the quirks cross the TAS (copper). The quirks of one event (blue) travel through the TAN (cooper) at $z\sim$ 140 m, which makes $\vec{L}_{tot}^\prime$ lifted again, while the quirks of the other event (red) bypass the TAN. The changes of $\vec{L}_{tot}^\prime$ when the quirk pairs inject into the concrete are different in two events because the impact of $\vec{F}_\text{ion}$ is turned on at different values of $\rho$ in two events. During the quirks travelling in around 100 meters of concrete and rock, $\vec{L}_{tot}^\prime$ of both events fluctuate mildly due to the Gaussian PDF of $F_i(v_i)$, which is consistent with our discussion above.

	\subsection{Signal efficiency}
	
	Solving the quirk EoM, we can test whether a quirk pair with given initial momenta could enter the tracker of FASER (2). 
	The odd tracks induced by the quirk pair can be identified easily in the FASER (2) tracker because they are very different from the tracks of SM particles. 
	The quirk pair satisfying $p_T(\mathcal{QQ})/|p(\mathcal{QQ})|>0.002$ can enter the FASER 2 tracker when its oscillation amplitude reaches the detector size $\sim \mathcal{O}(1)$ m (corresponds to $\Lambda \lesssim 100$ eV).

	\begin{figure}[thb]
		\begin{center}
\includegraphics[width=0.24\textwidth]{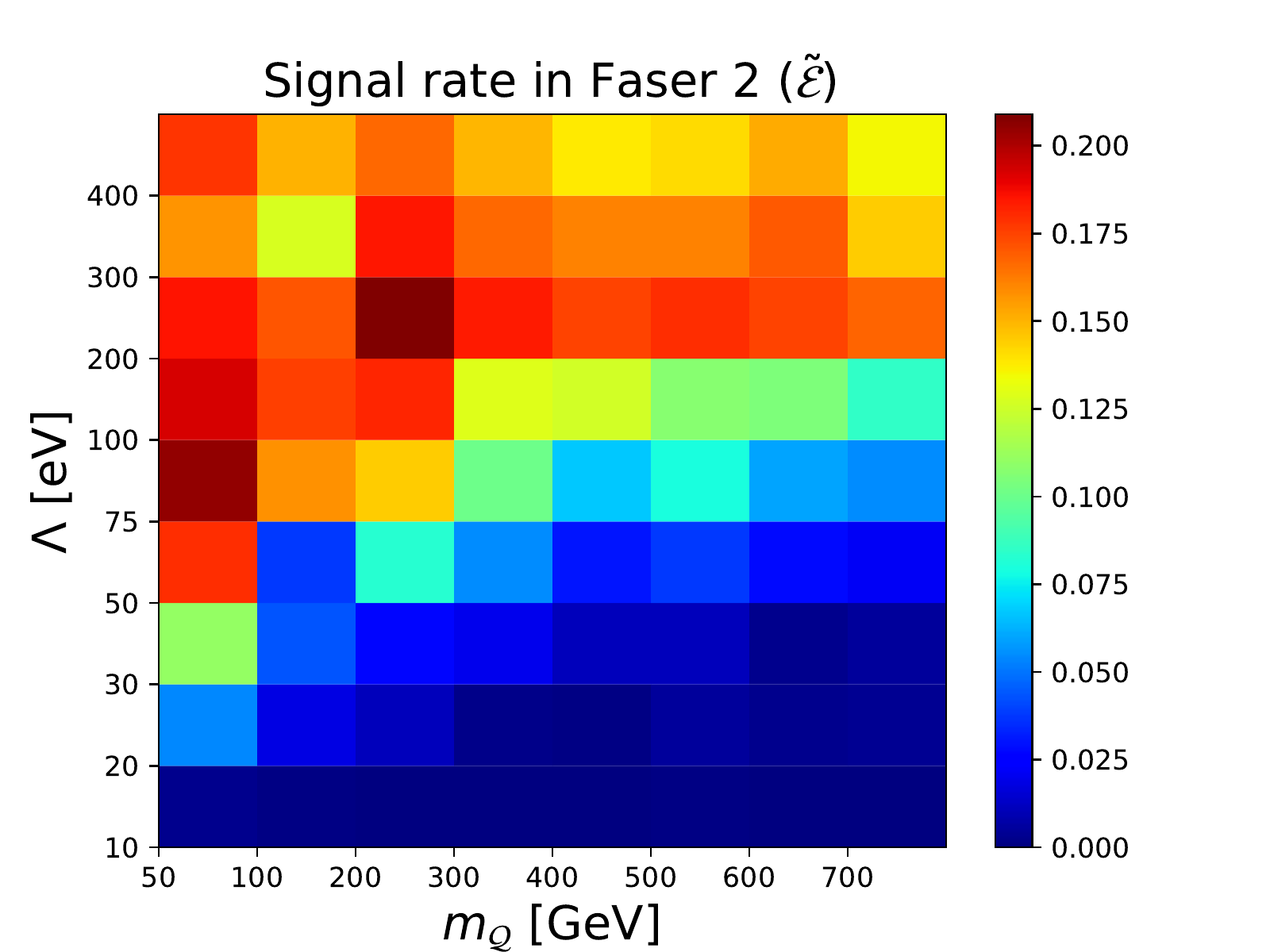}
\includegraphics[width=0.24\textwidth]{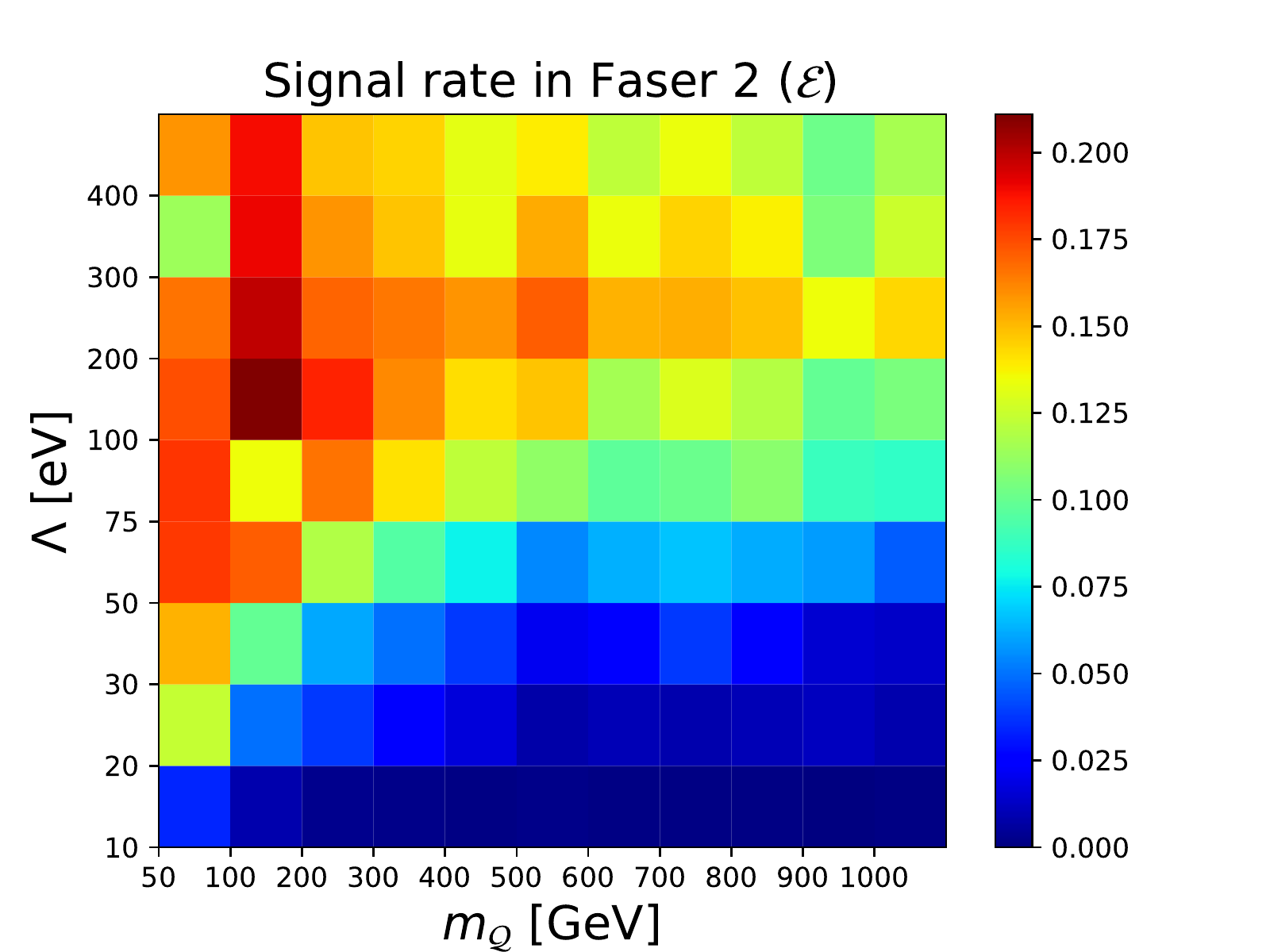} 
\includegraphics[width=0.24\textwidth]{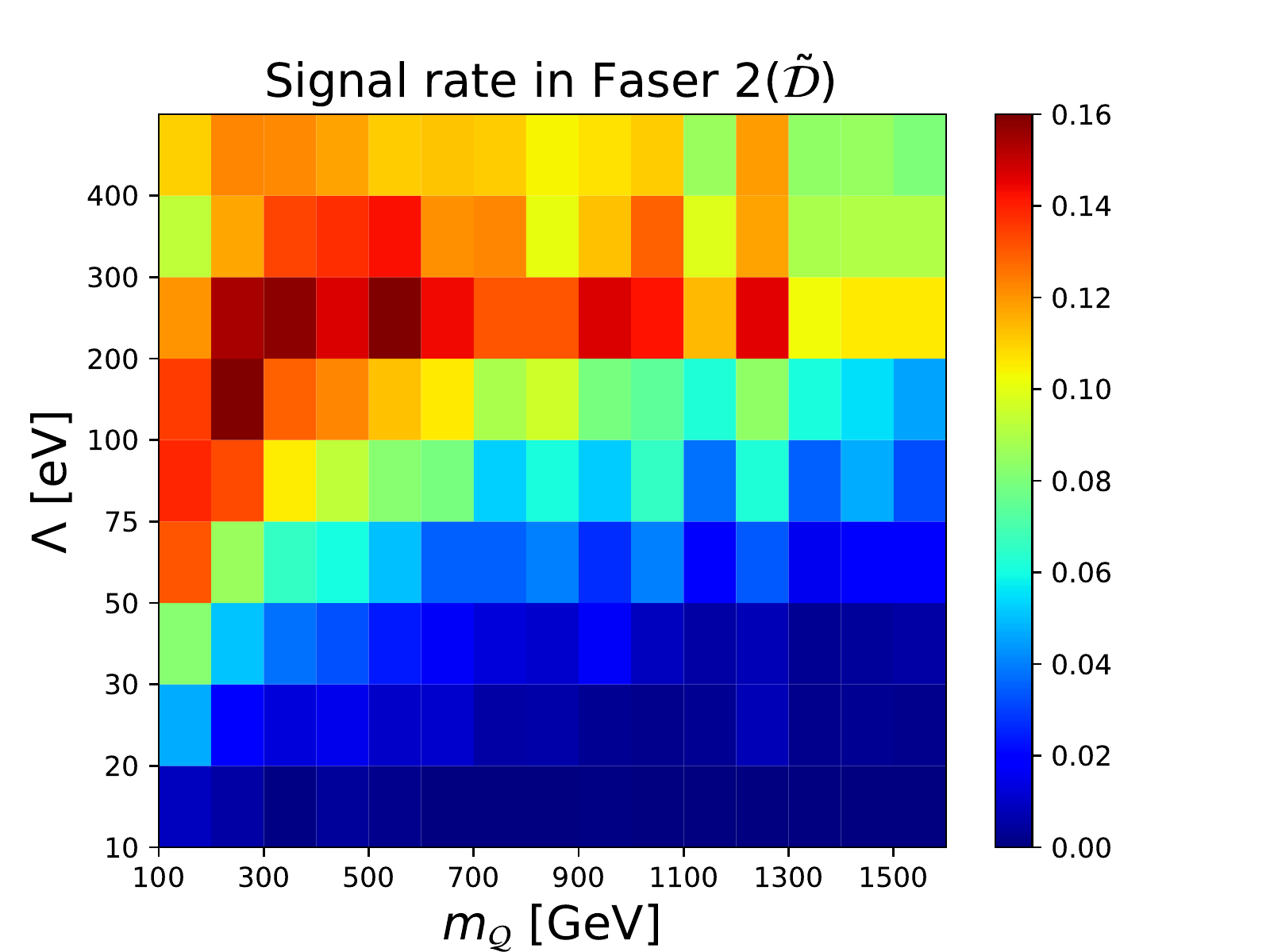}
\includegraphics[width=0.24\textwidth]{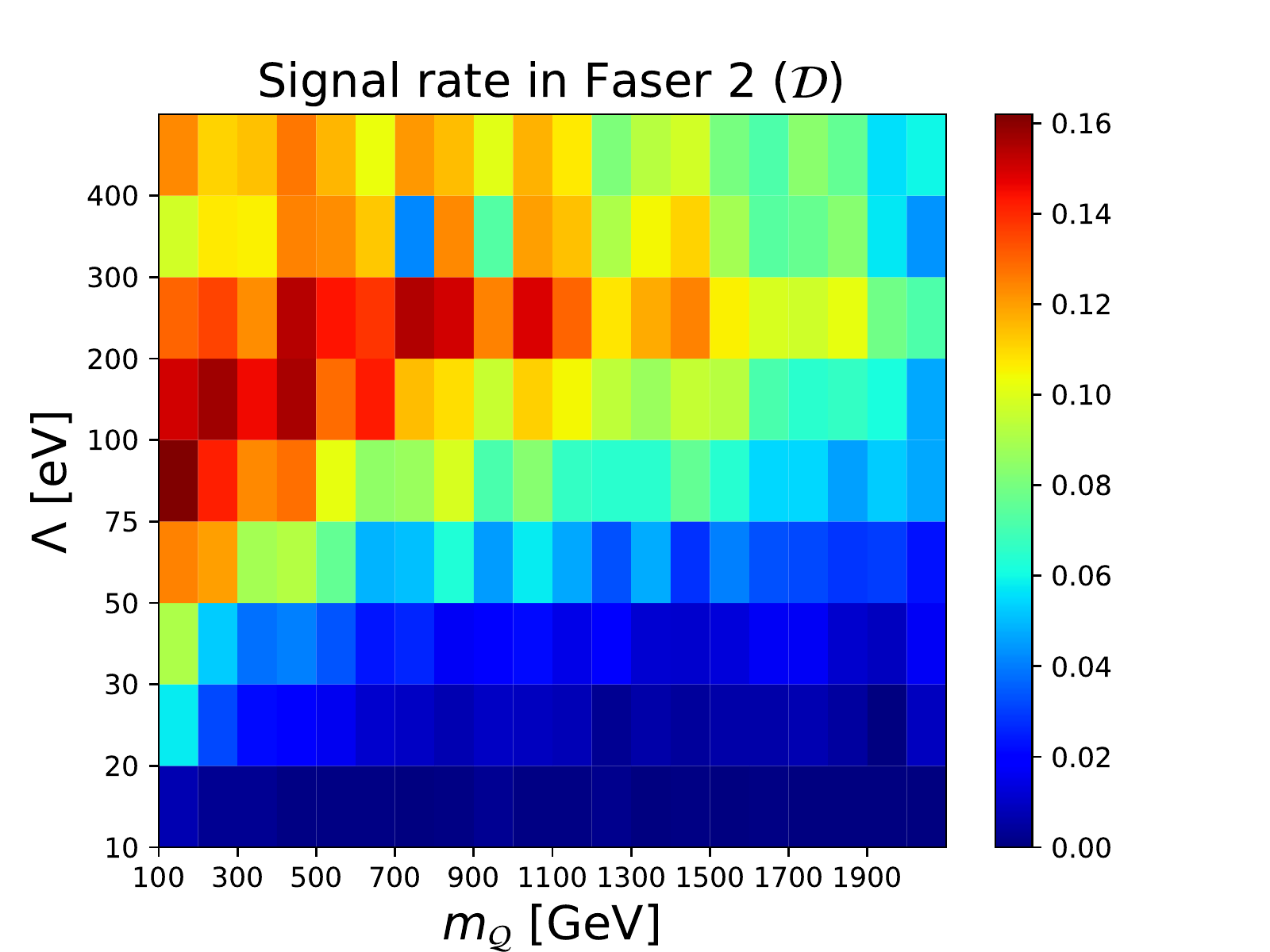}\\
\includegraphics[width=0.24\textwidth]{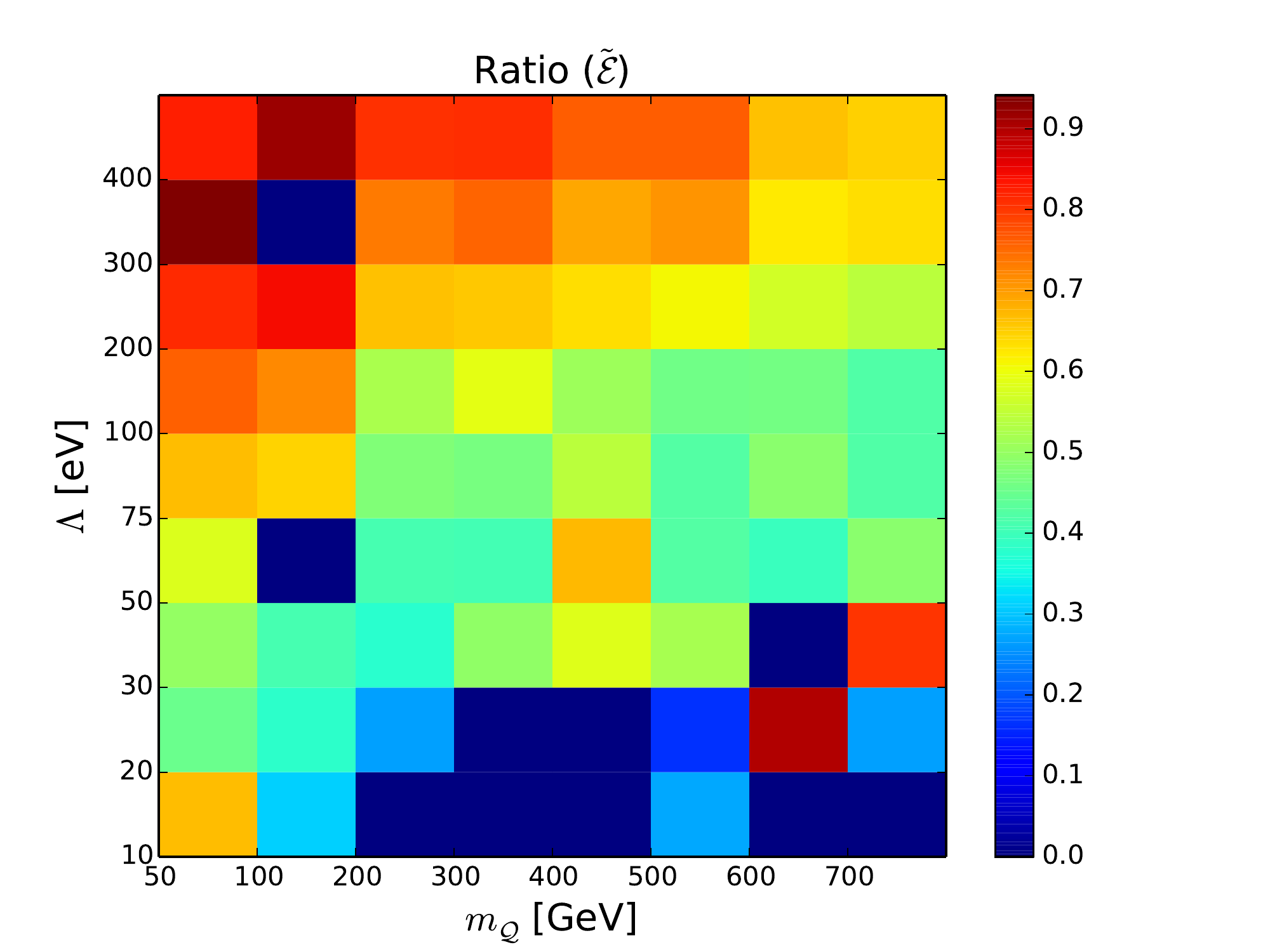}
\includegraphics[width=0.24\textwidth]{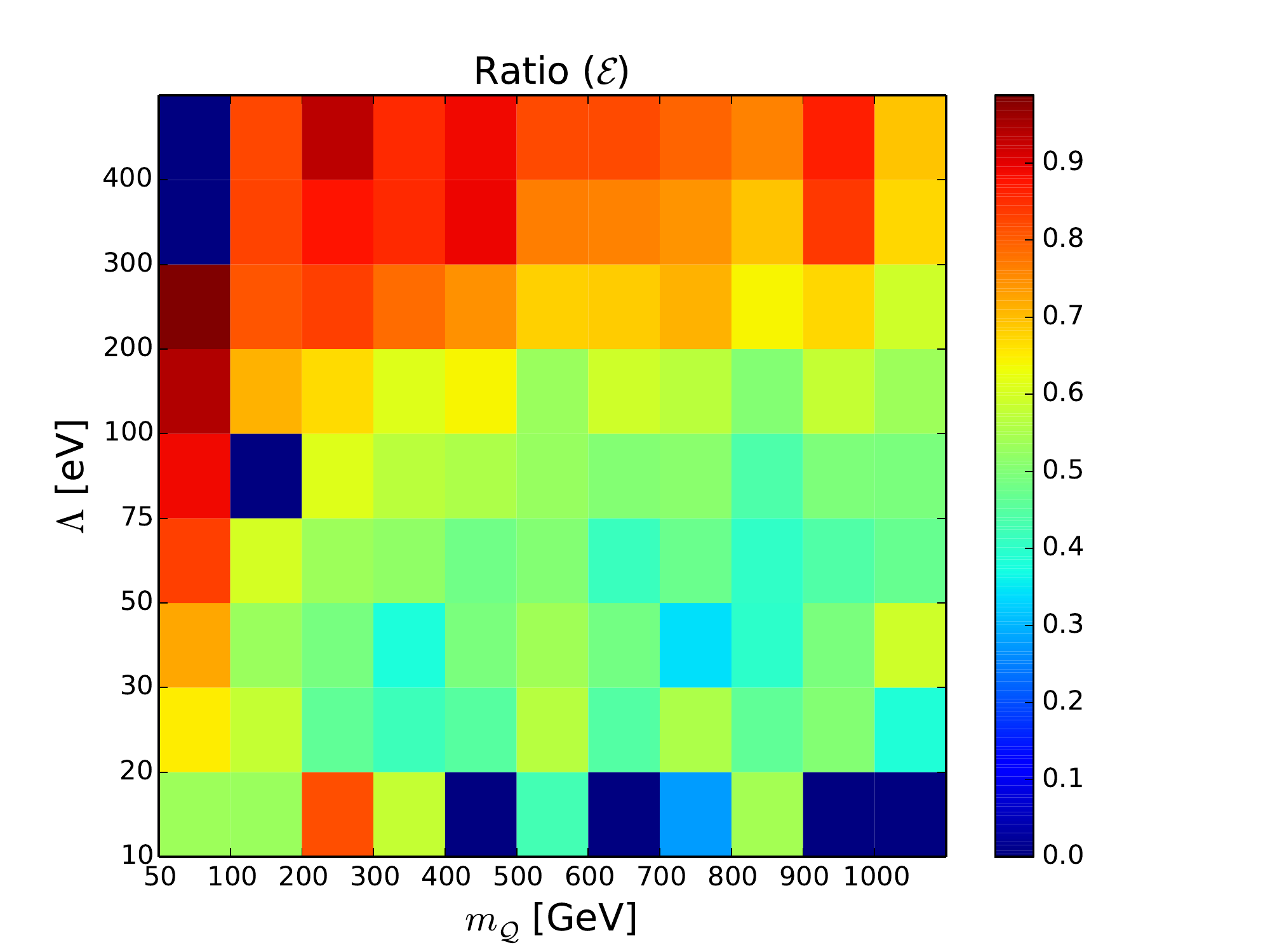} 
\includegraphics[width=0.24\textwidth]{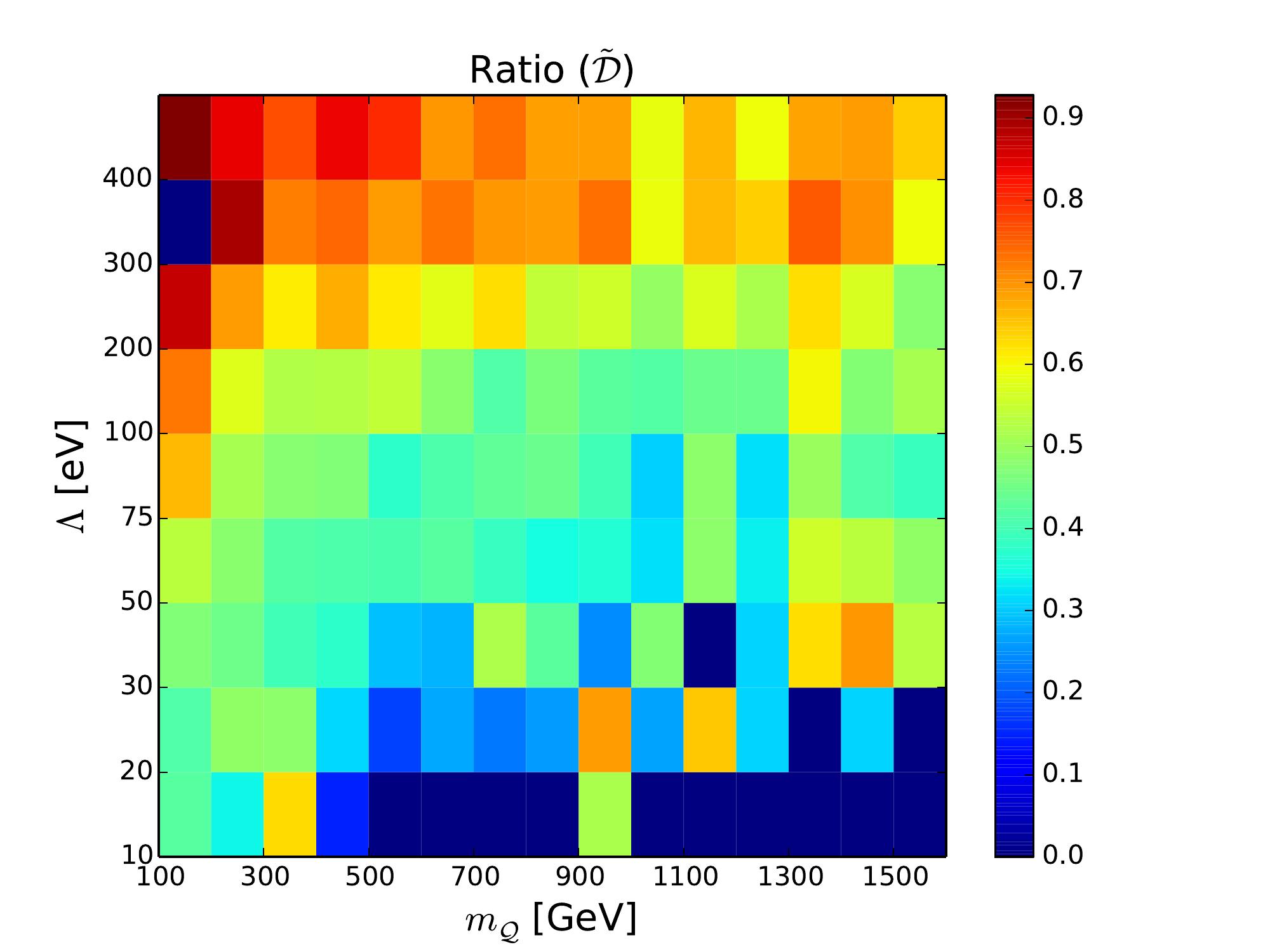}
\includegraphics[width=0.24\textwidth]{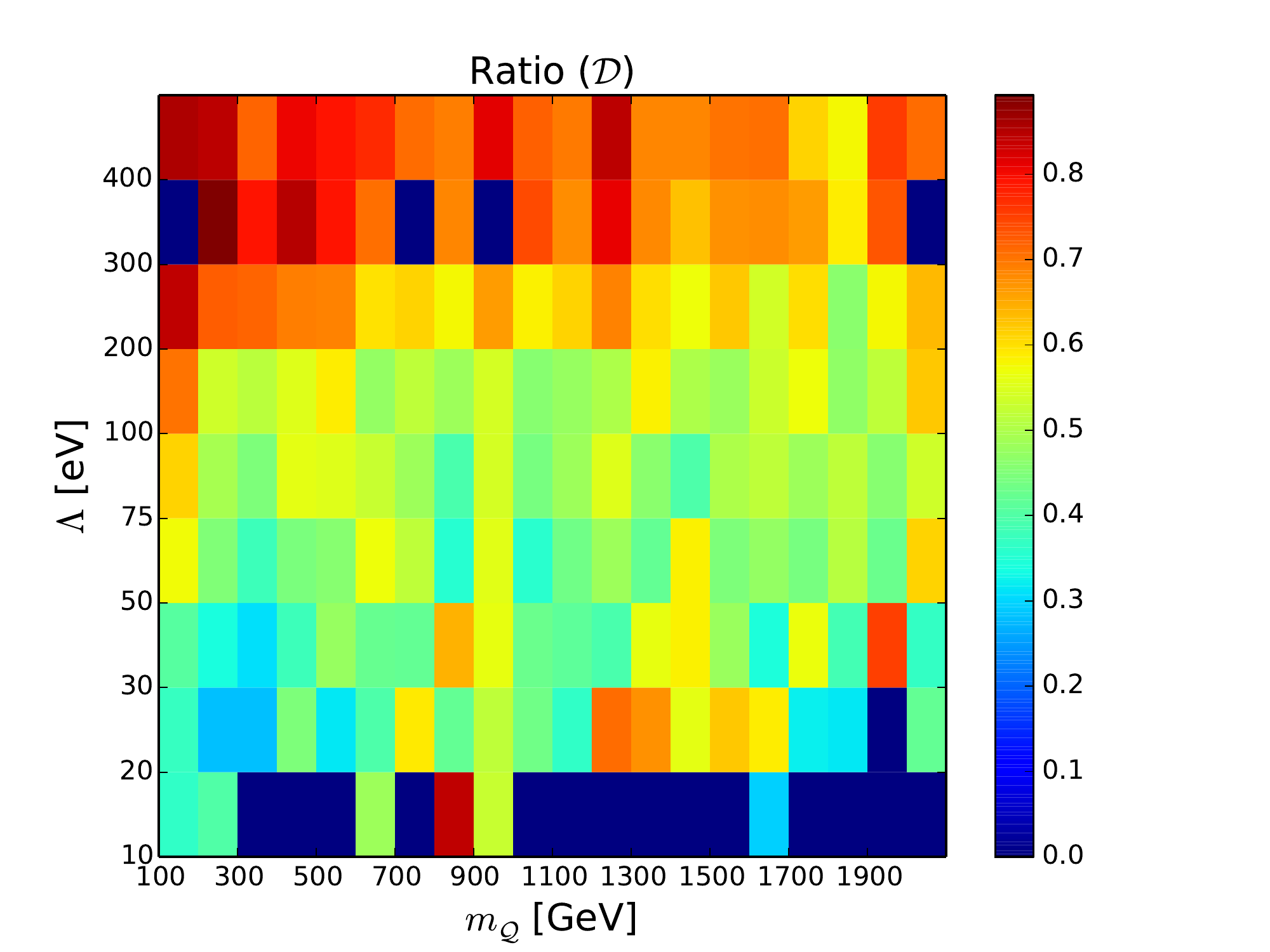}
		\end{center}
		\caption{ Upper panels: the fractions of quirk events (in event sample with $p_T(\mathcal{Q}\mathcal{Q})/ |p(\mathcal{Q}\mathcal{Q})|< 0.005)$ that have at least one quirk entering the FASER 2 tracker. Lower panels: among the events which can enter the FASER 2 tracker, the ratio between the number of events with $p_T(\mathcal{Q}\mathcal{Q})/ |p(\mathcal{Q}\mathcal{Q})|< 0.002$ and $p_T(\mathcal{Q}\mathcal{Q})/ |p(\mathcal{Q}\mathcal{Q})|< 0.005$ in initial state. Quirks with four different quantum numbers as given in Eq.~\ref{eq::qn1}-Eq.~\ref{eq::qn4} are considered.}
		\label{fig::tag0052}
	\end{figure}
	
	In the upper panels of Figure~\ref{fig::tag0052}, the fractions of quirk events (in the event sample which satisfies $p_T(\mathcal{Q}\mathcal{Q})/ |p(\mathcal{Q}\mathcal{Q})|< 0.005$) that have at least one quirk entering the FASER 2 tracker are shown on the $m_{\mathcal{Q}} - \Lambda$ plane. 
	For quirk production at the LHC, heavier quirk has lower velocity, thus suffers from stronger ionization force. 
	For a heavy quirk-pair system with initial momentum pointing to the FASER 2 tracker, the ionization force will continually change the moving direction and make the quirk pair system miss the FASER 2 tracker eventually.  What is more important, heavier quirk mass leads to a larger oscillation amplitude of the quirk pair, and thus makes the quirk pair more likely to bypass the FASER 2 tracker when  $p_T(\mathcal{Q}\mathcal{Q})/ |p(\mathcal{Q}\mathcal{Q})|$ is fixed.
	So, the signal efficiency becomes lower for heavier quirk.
	
	The dependence of the signal efficiencies on the confinement scale $\Lambda$ is more interesting. In cases with $|\vec{F}_s| \gg |\vec{F}_{\text{ion}}|$, the characteristic oscillation amplitude of the quirk pair in the lab frame can be expressed as~\cite{Li:2020aoq} 
	\begin{align}
		L=\frac{\mathcal{R}}{\rho}\ell_{c} ~,~\label{eq::LL}
	\end{align}
	where
	\begin{align}
		&\ell_{c}=2~\text{cm}(\sqrt{1+\rho^2} -1)
		\frac{m}{\text{[100~GeV]}}\frac{\text{[keV]}^2}{\Lambda^2}~, \\
		&\rho  =\sqrt{\frac{(E_1+E_2)^2-(\vec{P}_1+\vec{P}_2)^2}{4m^2}-1}~,\\
		&\mathcal{R}=\frac{|\vec{P}_1\times\vec{P}_2|}{m|\vec{P}_1+\vec{P}_2|}~.
	\end{align} 
	$\vec{P}_1$ ($E_1$) and $\vec{P}_2$ ($E_2$) are initial momenta (energies) of two quirks in the lab frame, and $\rho$ is same as that in Eq.~\ref{rcom}.
	The $\ell_{c}$ is half of the largest distance between the two quirks during the oscillation in the CoM frame~\cite{Kang:2008ea}.
	The $L$ corresponds to the length of the projection of $\ell_{c}$ onto the plane perpendicular to $\vec{P}_1+\vec{P}_2$, which stands for half the width of the belt that can cover the trajectories of the two quirks.  
	When the $\Lambda$ is sizable such that the oscillation amplitude $L \ll \mathcal{O}(1)~\text{m}$, the signal efficiency at FASER 2 is given by the initial moving direction of the quirk-pair system, {\it i.e.} the direction of $\vec{P}_1+\vec{P}_2$. 
	Since the dimension of the FASER 2 tracking plane is taken as $2\times 2~\text{m}^2$ and the distance from the IP to FASER 2 is 480 m, only the quirk pairs satisfying $p_T(\mathcal{QQ})/|p(\mathcal{QQ})|<0.002$ can enter the FASER 2 tracker. 
	On the other hand,  in the small $\Lambda$ region where $L \gtrsim \mathcal{O}(10)~\text{m}$, many quirk events will bypass the FASER 2 tracker even though the $p_T(\mathcal{QQ})/|p(\mathcal{QQ})|<0.002$ condition is fulfilled, leading to very low signal efficiency. 
	The signal efficiency is highest in the moderate $\Lambda$ region where $L\sim\mathcal{O}(1)~\text{m}$. In this region, beside the events with $p_T(\mathcal{QQ})/|p(\mathcal{QQ})|<0.002$, others with $p_T(\mathcal{QQ})/|p(\mathcal{QQ})| \gtrsim 0.002$ will also enter the FASER 2 tracker due to the sizable oscillation amplitudes.

	The above point is also illustrated in the lower panels of Figure~\ref{fig::tag0052}, which show the ratio between the number of events with initial $p_T(\mathcal{Q}\mathcal{Q})/ |p(\mathcal{Q}\mathcal{Q})|< 0.002$ and that with initial $p_T(\mathcal{Q}\mathcal{Q})/ |p(\mathcal{Q}\mathcal{Q})|< 0.005$ among the events that can reach the FASER 2 tracker.   
	For large $\Lambda$, the ratio is close to one, which means only those events with $p_T(\mathcal{Q}\mathcal{Q})/ |p(\mathcal{Q}\mathcal{Q})|< 0.002$ can reach the FASER 2 tracker. And the ratio is lower for smaller $\Lambda$, {\it i.e.} some events with $p_T(\mathcal{QQ})/|p(\mathcal{QQ})| \gtrsim 0.002$ are secured by the large oscillation amplitudes. Note that the results for $\Lambda \le 30$ eV can not be trusted because there are only a few events that can reach the FASER 2 tracker and the fluctuation is huge in our simulation. 

	
	\subsection{Infracolor glueball and electromagnetic radiations}
	
	In solving the EoM of quirks, the effects of infracolor glueball and electromanetic radiations are not considered. 
	As has been discussed in Ref.~\cite{Kang:2008ea}, both effects can be parameterized by assuming that the energy radiated in a crossing time $t_p$ is $\Delta E$, with $\Delta E_{IC} = \epsilon \Lambda_{IC}$ and $\Delta E_{EM} = \alpha_{EM} / t_p$. The $\epsilon$ corresponds to the probability of infracolor glueball emission during each crossing, which is $\sim \mathcal{O}(0.1)$~\cite{Evans:2018jmd}. The $\Delta E_{EM}$ is estimated from the Larmor formula. 
	In cases with $|\vec{F}_s| \gg |\vec{F}_{\text{ion}}|$, the analytical solution for the quirk EoM studied in Ref.~\cite{Li:2020aoq} shows that the crossing time of the quirk pair is
	
	\begin{align}
		t_{p}=0.132 \sqrt{\left( \frac{E_1+E_2}{2m}\right) ^2-\frac{1}{1-\beta^2}} \frac{m}{[100~\mathrm{GeV}]} \frac{[\mathrm{keV}]^{2}}{\Lambda^{2}}[\mathrm{ns}]~, \label{eq::tp}
	\end{align}
	corresponding to a travel distance of
	\begin{align}
		d_{p}=0.3\beta\frac{t_p}{[\mathrm{ns}]}[\mathrm{m}],
	\end{align} 
	where $\beta$ ($\equiv\frac{|\vec{P}_1+\vec{P}_2|}{E_1+E_2}$) is the velocity of the quirk-pair system in the lab frame. 
	For the quirk pair traveling from the ATLAS IP to the FASER (2) detector, the typical number of crossing can be calculated as 
	\begin{align}
		N_{\text{crossing}} = \frac{480~[m]}{d_p} = \frac{120}{\beta \sqrt{(\frac{E_1+E_2}{2m})^2 - \frac{1}{1-\beta^2}}} \frac{[100~\text{GeV}]}{m} \frac{\Lambda^2}{ [\text{keV}]^2}~.~
	\end{align}
	Based on the events that contain at least one quirk reaching the FASER (2) detector, we find that the $N_{\text{crossing}}$ depends quite mildly on the quirk mass and it can be estimated as $N_{\text{crossing}} \sim 20 \times \frac{\Lambda^2}{[100~\text{eV}]^2}$. 
	
	Eventually, the total energy losses due to infracolor glueball and electromagnetic radiations can be estimated:
	\begin{align}
		E_{IC} & \sim  2 \times \frac{\Lambda^2}{ [100~\text{eV}]^2} \times \Lambda~,~ \\ 
		E_{EM} & \sim  1.7 \times 10^{-7} \times \alpha_{EM} \times \frac{\Lambda^4}{[100~\text{eV}]^4} ~\text{[eV]} ~.~
	\end{align}
	Both values are much smaller than the typical kinetic energy of quirk at the LHC. So it is safe to simply ignore the effects of infracolor glueball and electromagnetic radiations in our simulation.

	\subsection{The FASER (2) sensitivity}
	
	The total number of quirk events in FASER (2) can be calculated by 
	\begin{align}
		N_{\text{sig}} = \sigma \times \epsilon_{\rm fid} \times \epsilon_{0.005} \times \mathcal{L},
	\end{align}
	where the quirk production cross section $\sigma$~\footnote{The colored quirks $\tilde{\mathcal{D}}$ and $\mathcal{D}$ will hadronize into quirk-quark bound states, and the probability for those final states to have $\pm 1$ charges is roughly 30\%. So, the factor 0.3 is multiplied on their cross sections when estimating the number of signal events.} and the fiducial efficiency $\epsilon_{\rm fid}$ are illustrated in Figure~\ref{fig::xsecs}, and the $\epsilon_{\rm fid}$ corresponds to the efficiency of selecting events with $p_T(\mathcal{Q}\mathcal{Q})/ |p(\mathcal{Q}\mathcal{Q})|< 0.005$ in quirk pair production. Events with $p_T(\mathcal{Q}\mathcal{Q})/ |p(\mathcal{Q}\mathcal{Q})|> 0.005$ and reaching FASER (2) are not counted because events in this kinematic region have quite low efficiency and solving the quirk EoM for them will take too much time. Including those events will only improve the FASER (2) sensitivity a little bit. 
	The signal efficiency $\epsilon_{0.005}$ is shown in the upper panels of Figure~\ref{fig::tag0052}. And the integrated luminosity $\mathcal{L}$ is taken as $150$ and $3000$ fb$^{-1}$ for FASER and FASER 2, respectively. 
	
\begin{figure}[thb]
	\begin{center}
		\includegraphics[width=1.0\textwidth]{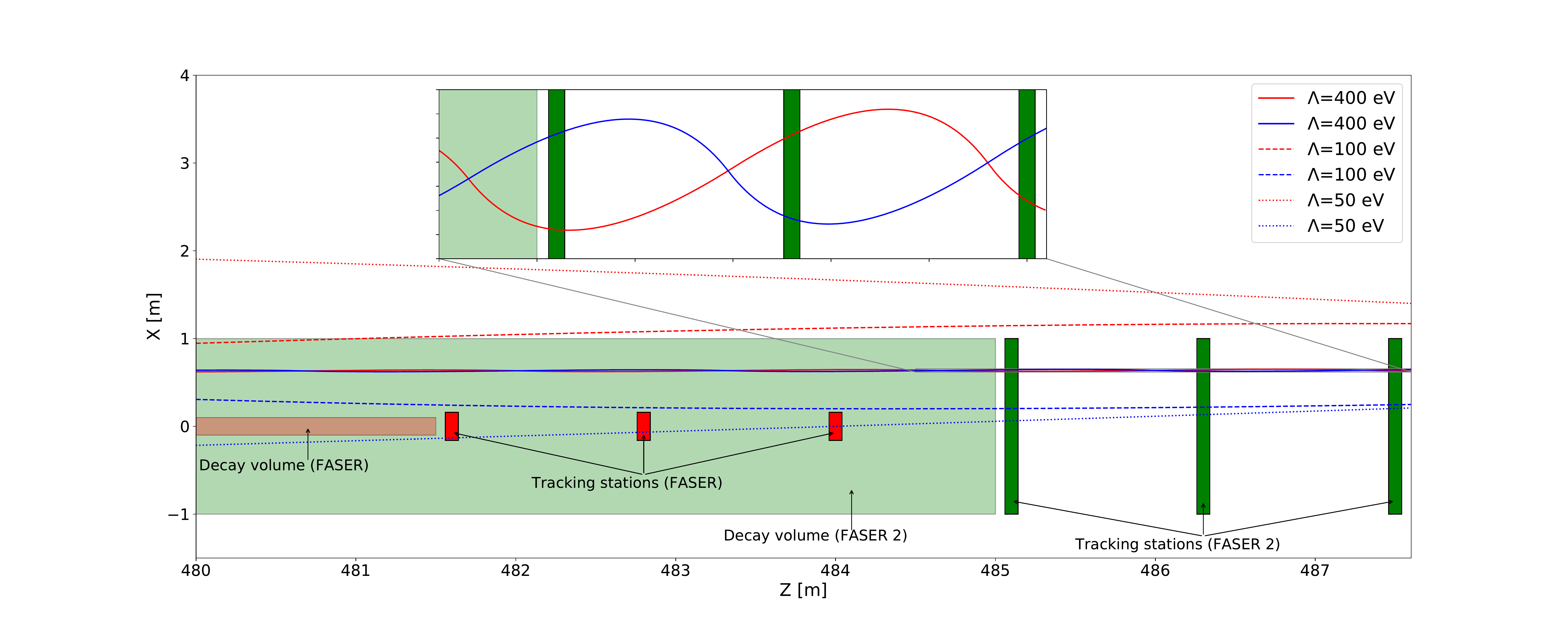}
	\end{center}
	\caption{The quirk trajectories in the FASER and FASER 2 detectors. The quirk initial momenta and mass are the same as those in Figure~\ref{fig::quirktrajectory}. Three different confinement scales $\Lambda=50~\text{eV}, 100~\text{eV}, 400~\text{eV}$ are considered for illustration and two trajectories with the same line style correspond to a quirk pair with fixed $\Lambda$. The red (green) and light red (green) regions indicate the decay volume and three tracking stations of FASER (2), respectively. }
	\label{fig::quirktrajectorytracker}
\end{figure}
	
The dominant source of background is the radiative processes associated with muons coming from the ATLAS interaction point. The expected flux of muon with energy greater than 100 GeV at the FASER (2) location can reach 0.2 cm$^{-2}$s$^{-1}$~\cite{FASER:2018ceo,FASER:2018eoc,FASER:2018bac}, for the LHC Run 3. In neutral long-lived particle searches, this background can be highly suppressed by using a charged particle veto layer at the front of the detector. As most of the quirks are entering FASER (2) from the front, the charged particle veto should not be applied in the quirk search. 
	However, the tracks of quirks have unique features that can be used to distinguish them from the tracks of high energy muons. 
	In Figure~\ref{fig::quirktrajectorytracker}, we illustrate the quirk trajectory inside the FASER (2) detector. The same quirk event as in Figure~\ref{fig::quirktrajectory} is used here. 
	First of all, the typical oscillation amplitude of quirk pair for quirk mass $\sim \mathcal{O}(100)$ GeV and confinement scale $\Lambda \in [100,1000]$ eV is $\gtrsim \mathcal{O}(1)$ cm, and the hit position resolution of the FASER (2) tracker is much smaller (300 $\mu$m). So, quirk tracks are characterized by a pair of hits on each tracking plane. 
	Secondly, FASER (2) contains 9 tracking planes in 3 tracking stations. The hits of quirks on those tracking planes can not be reconstructed as helical tracks (but the hits of both quirks lie approximately on a single plane~\cite{Knapen:2017kly}).
	Thirdly, for the case when both hits from the quirk pair are measured on tracking planes, the centers of the hit pair will nearly lie on a straight line and point to the ATLAS IP. Note that this feature is also very useful to suppress the muon background from cosmic ray~\cite{Hill:1389902}. 
	Fourthly, the quirk mass is much heavier than muon, the typical velocity of quirk is much smaller than that of background muons. The ionization pattern~\cite{Li:2019wce} in silicon strip and the timing information will also help distinguishing quirks from muons. 
	Quantitive discussions on more quirk features which are useful for measuring quirk properties will be given in the next section. 
	Designing specific cuts to separate quirk tracks from muon tracks requires detailed simulation of detector effects, which is beyond the scope of the current work. 
	In the following, we simply assume that the background can be suppressed to a negligible level by using the features discussed above.  
}

	\begin{figure}[thb]
		\begin{center}
			\includegraphics[width=0.45\textwidth]{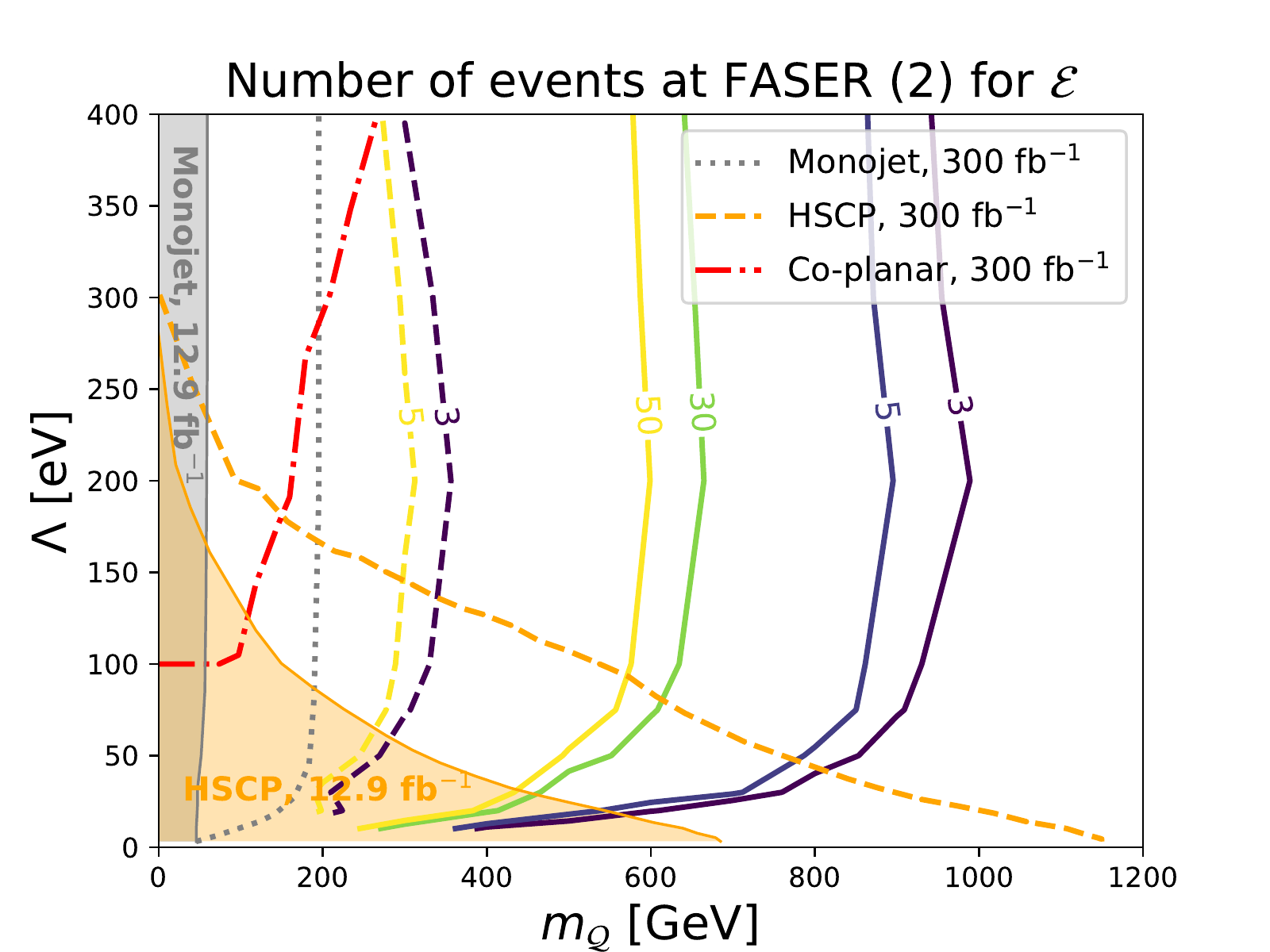}
			\includegraphics[width=0.45\textwidth]{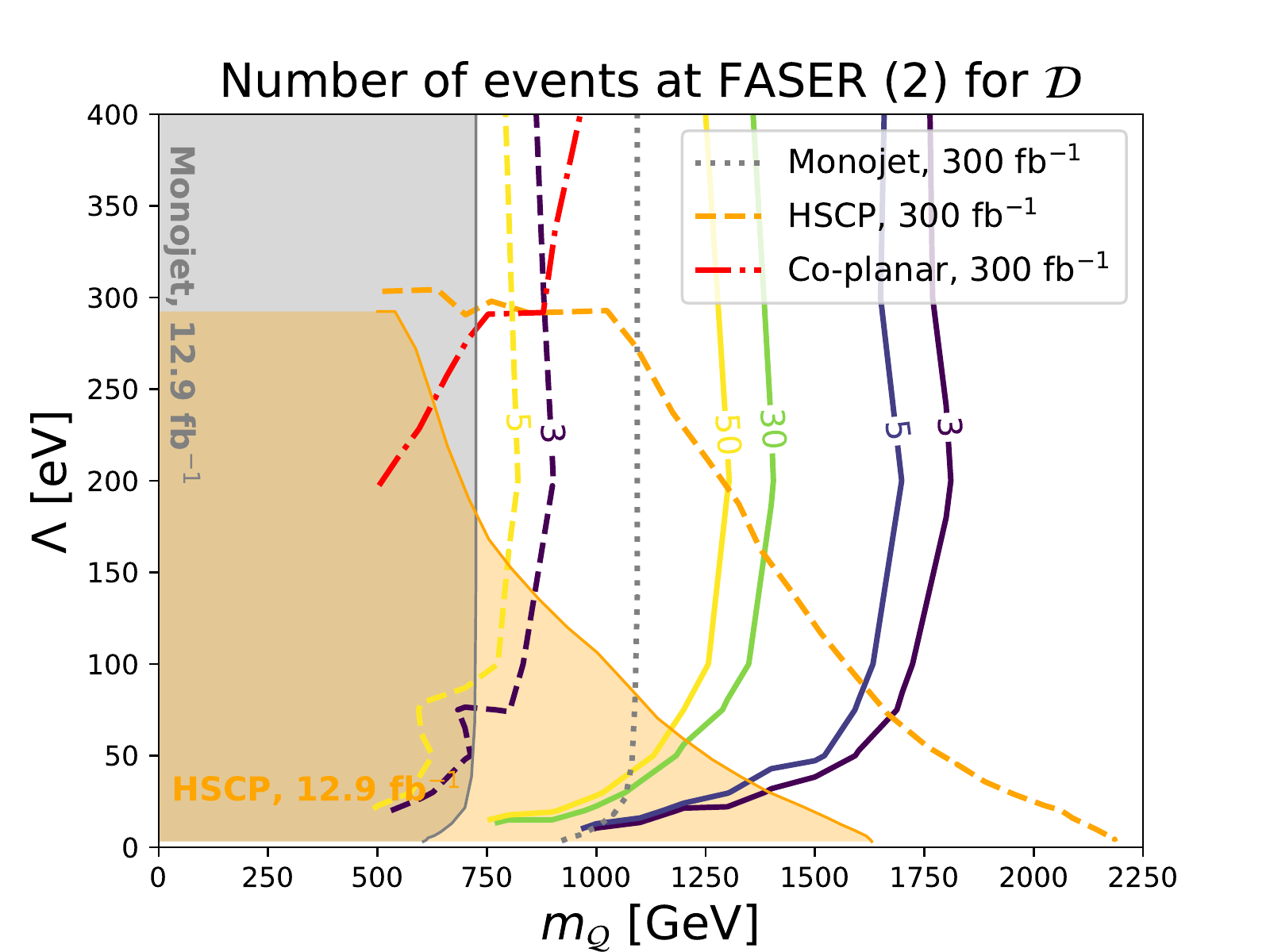} \\
			\includegraphics[width=0.45\textwidth]{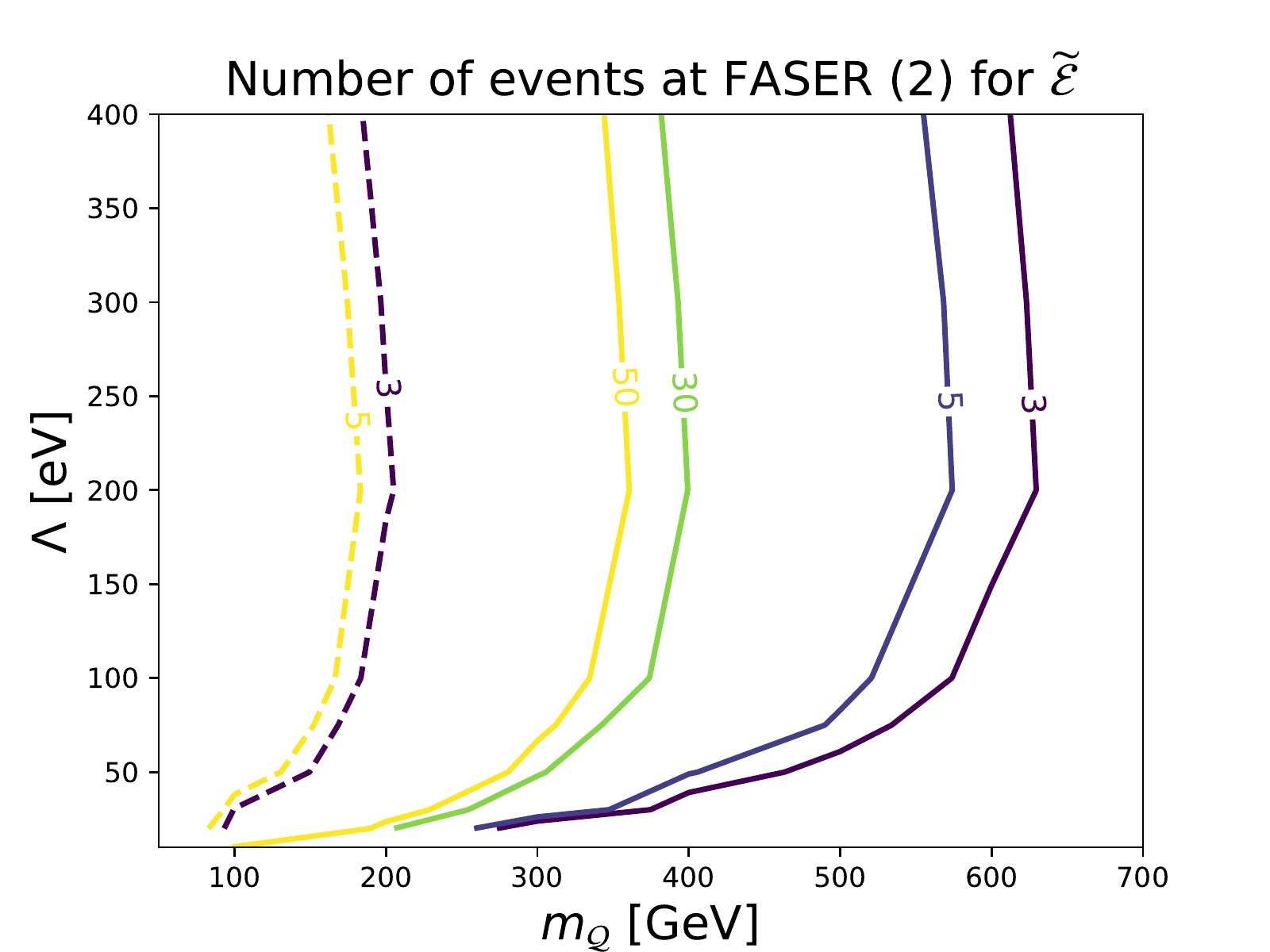}
			\includegraphics[width=0.45\textwidth]{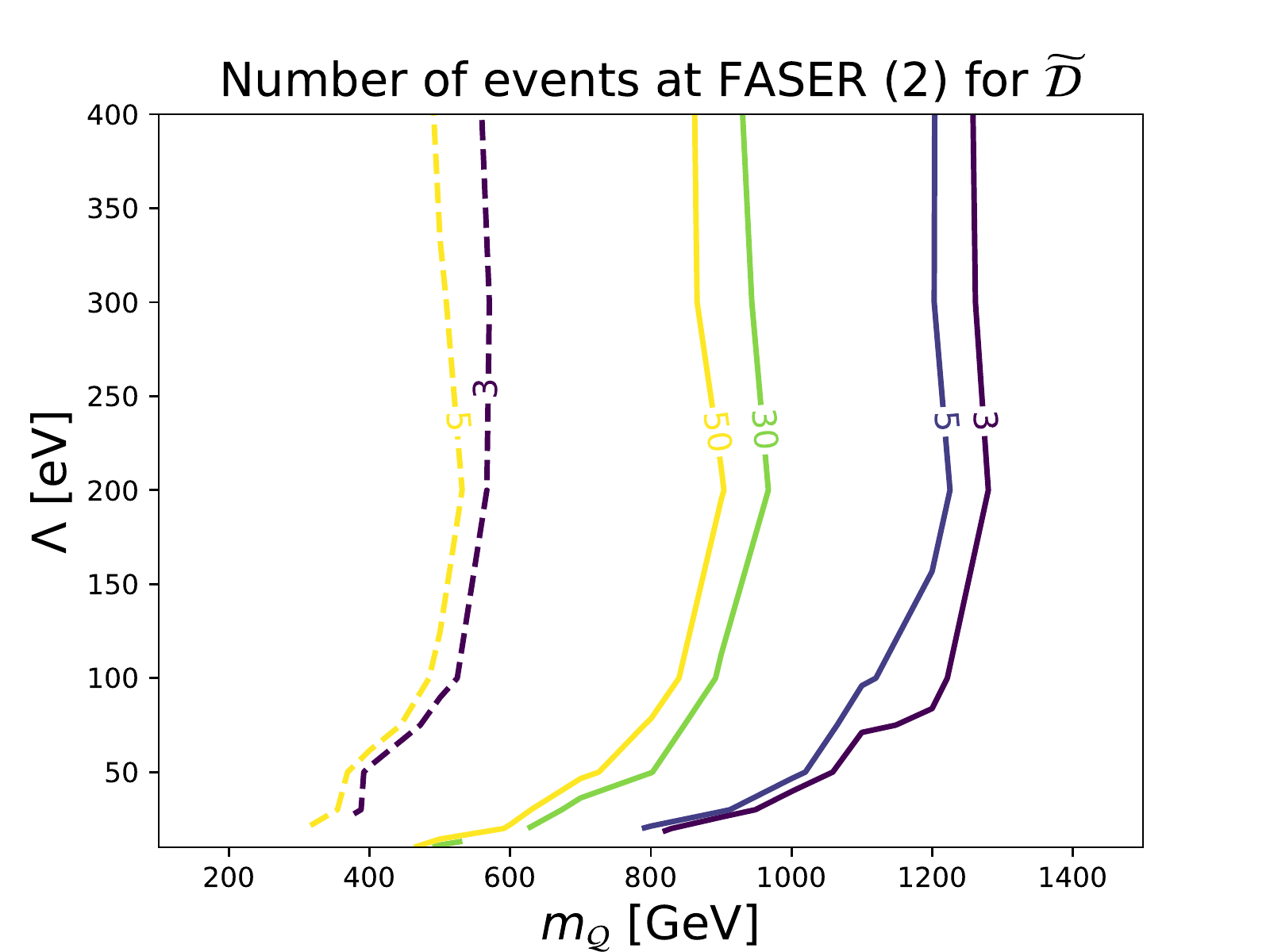} 
		\end{center}
		\caption{Contours of the number of quirk events that can reach the FASER (2) tracker in the $m_{\mathcal{Q}}$ versus $\Lambda$ plane for an integrated luminosity of 150 (3000) fb$^{-1}$. 
			The dashed (solid) ones correspond to the event numbers at FASER (2).
			For two fermionic quirks ($\mathcal{E}$ and $\mathcal{D}$), the projected bounds from Heavy Stable Charged Particle (HSCP) search~\cite{Farina:2017cts}, mono-jet search~\cite{Farina:2017cts}, and coplanar search~\cite{Knapen:2017kly} (the exclusion limits are taken) are shown by orange dashed line, grey dotted curves, and red dash-dotted line, respectively. Moreover, the existing bounds from the CMS HSCP search~\cite{CMS-PAS-EXO-16-036} and ATLAS monojet search~\cite{CMS:2016xbb} are shown by grey and orange shaded regions. Those bounds are extracted from Ref.~\cite{Knapen:2017kly}. }
		\label{fig::exclusion}
	\end{figure}
	
	{In Figure~\ref{fig::exclusion}, we show the number of quirk events at FASER and FASER 2 for an integrated luminosity of 150 fb$^{-1}$ and 3000 fb$^{-1}$, respectively. 
	The numbers are not very sensitive to the confinement scale $\Lambda$, as long as it is not too small, {\it i.e.} below 100 eV. 
	According to the discussions above, the unique features of quirk tracks can be used to highly suppress the background in the FASER (2) detector. 
	Given negligible background, the 3 events contours correspond to 2-$\sigma$ exclusion limits and the 5 events contours correspond to discovery prospects. 
	As a result, we can conclude that FASER 2 (FASER) will be able to exclude the $\mathcal{E}$, $\mathcal{D}$, $\tilde{\mathcal{E}}$ and $\tilde{\mathcal{D}}$ quirks with mass below 990 (360) GeV, 1800 (900) GeV, 630 (200) GeV and 1280 (570) GeV, respectively, with an integrated luminosity of 3000 (150) fb$^{-1}$ when $\Lambda \gtrsim \mathcal{O}(100)$ eV.
	The bounds on scalar quirks are much weaker than those on fermionic quirks when the gauge representations are the same because of their smaller production rates and lower signal efficiencies. FASER 2 is much more sensitive to the quirk signal than FASER because of the increased integrated luminosity as well as the larger tracking plane.
	For comparison, the projected bounds from the HSCP search, the mono-jet search, and the coplanar search as provided in Ref.~\cite{Knapen:2017kly} for fermionic quirks are shown as well. 
	For $\Lambda \lesssim 50$ eV, the HSCP search is most sensitive. FASER 2 is much more sensitive than other searches when $\Lambda \gtrsim \mathcal{O}(100)$ eV. 
	For the color neutral quirk $\mathcal{E}$, FASER behaves better than other searches when $\Lambda \gtrsim 150$ eV.
	
	\section{Features of the quirk signal}
	
	In this section, we discuss a few features of the quirk signals at the FASER (2) detector which can be used to separate them from possible backgrounds as well as resolve the quirk model parameters if a discovery can be made.

	\begin{figure}[thb]
		\begin{center}
\includegraphics[width=0.23\textwidth]{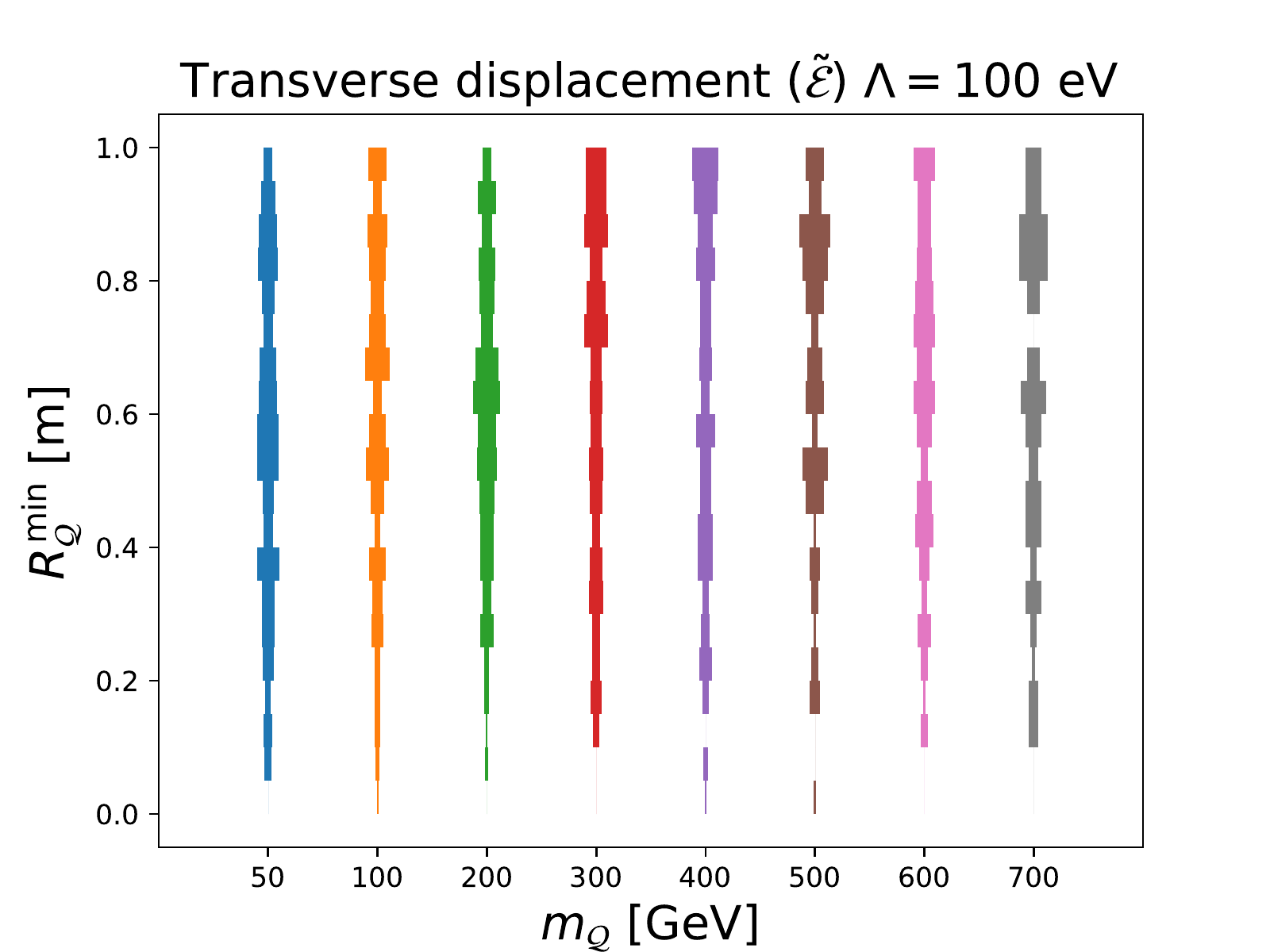}
\includegraphics[width=0.23\textwidth]{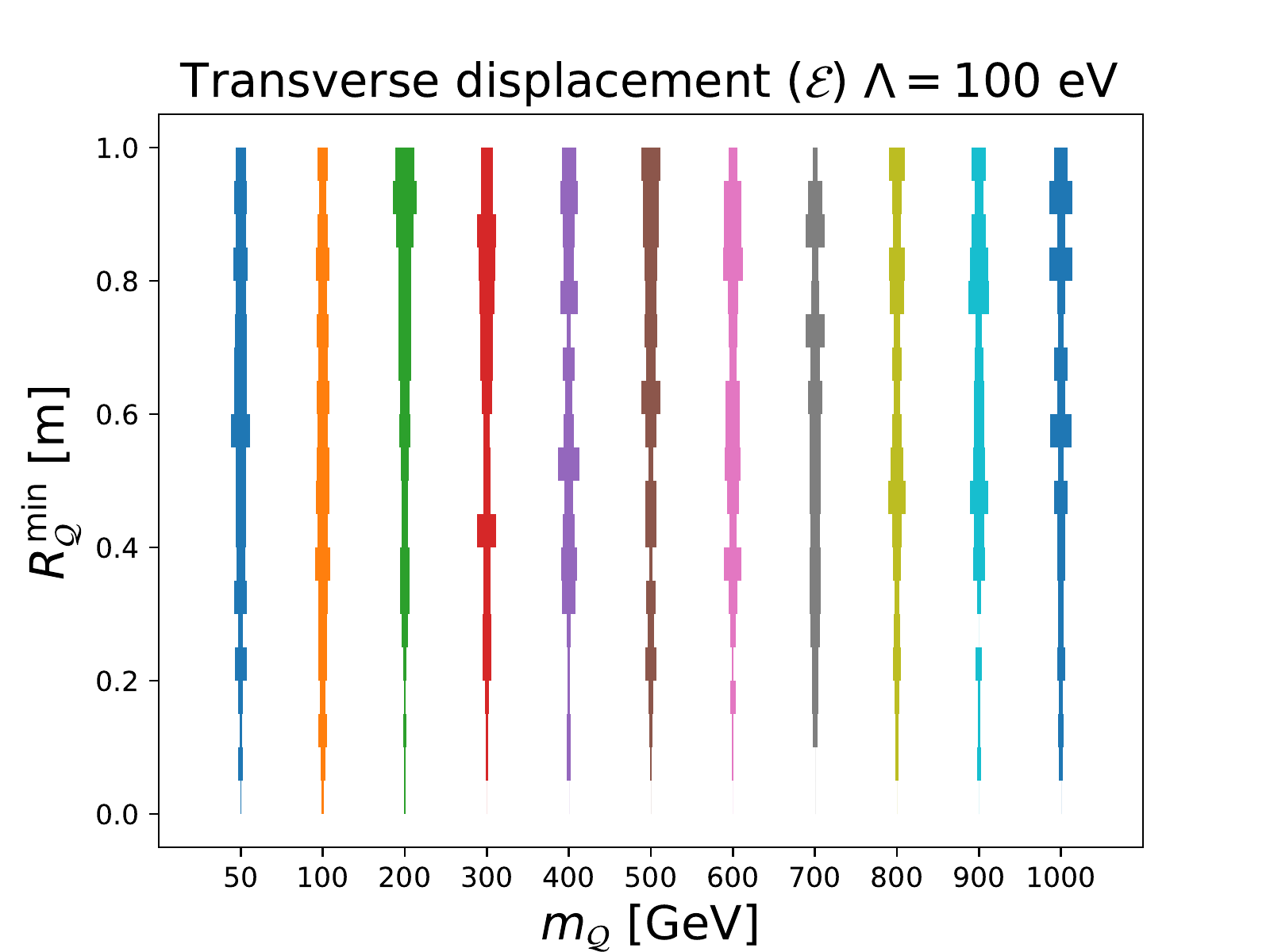} 
\includegraphics[width=0.23\textwidth]{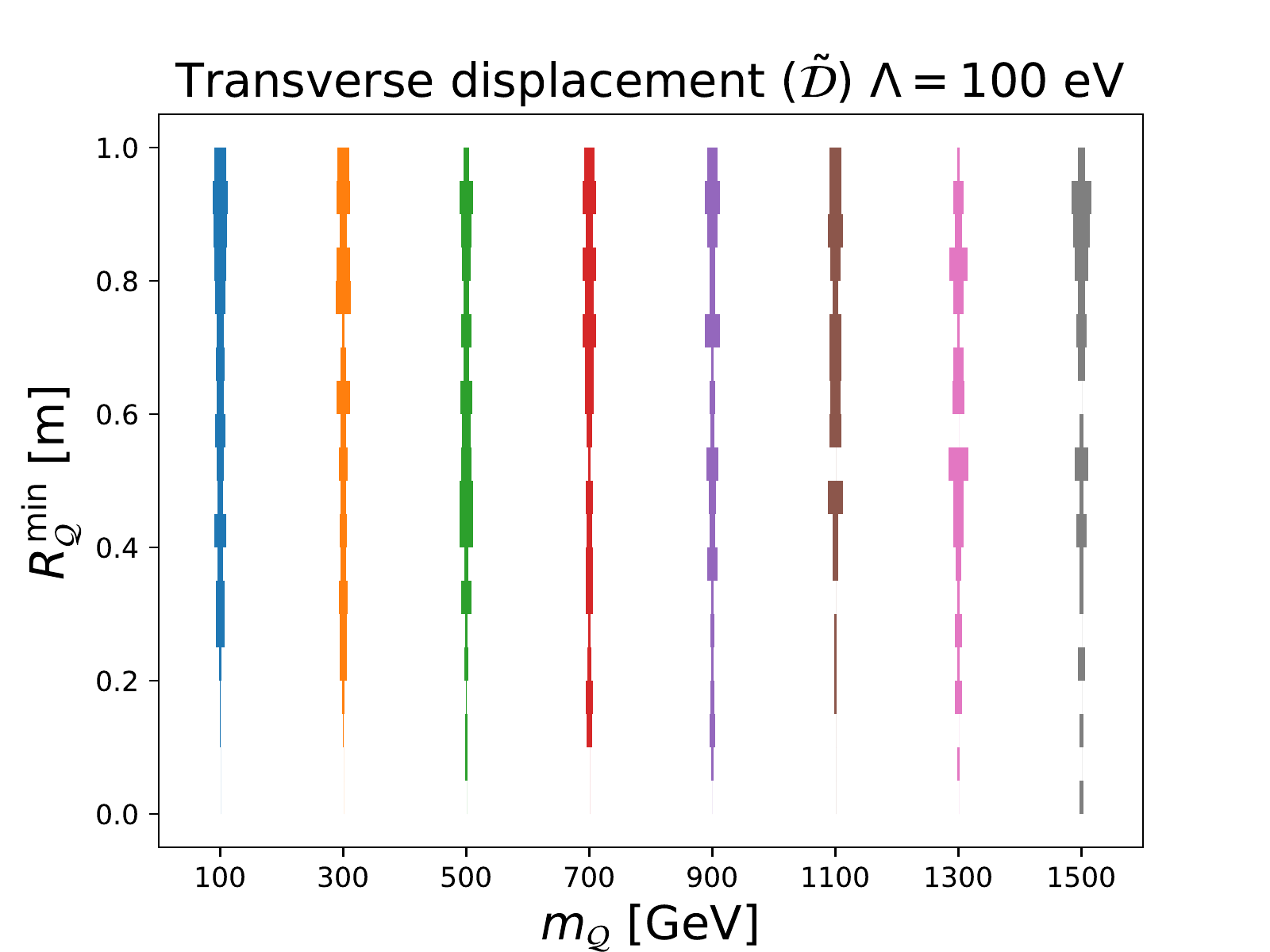}
\includegraphics[width=0.23\textwidth]{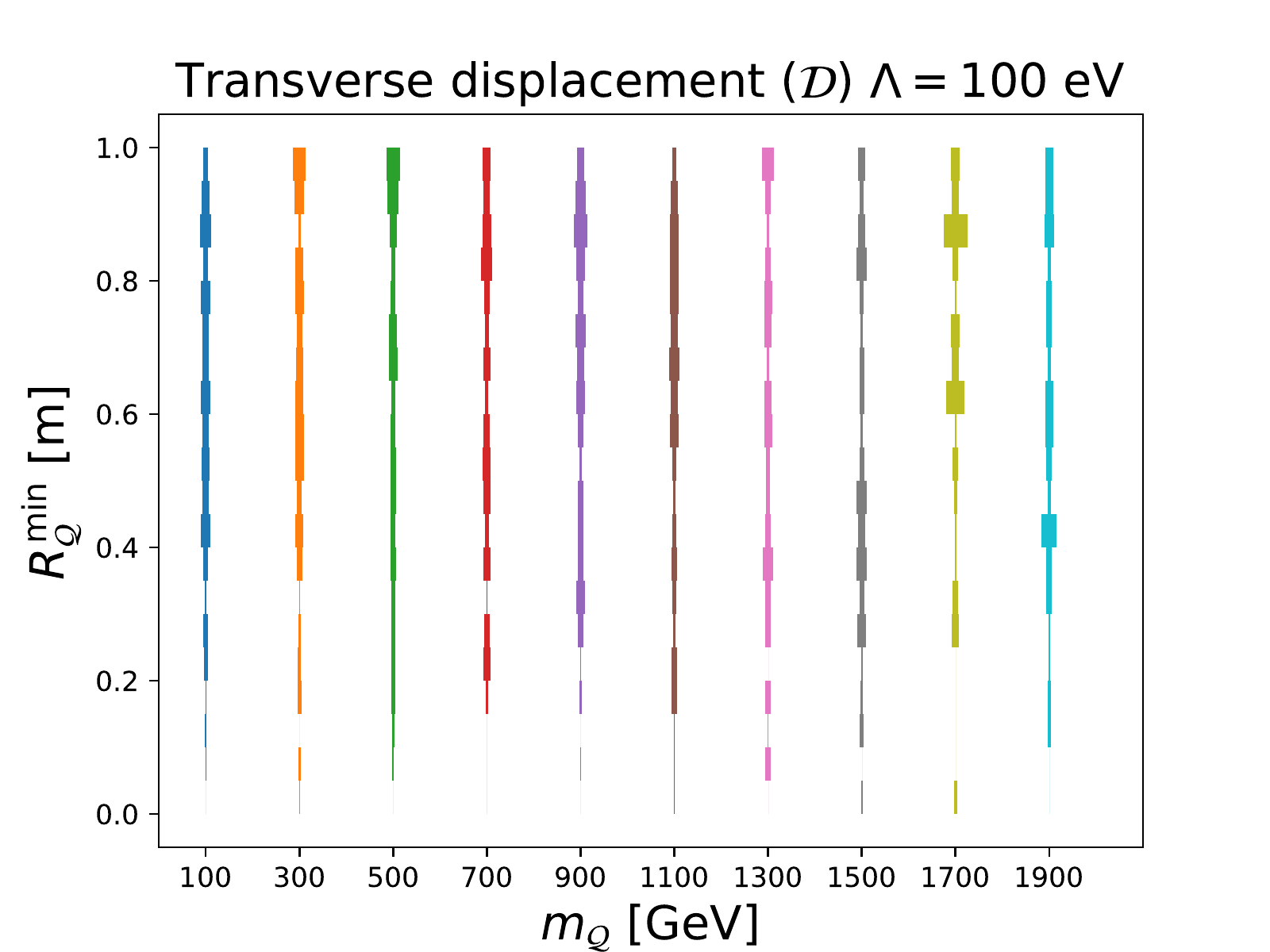} \\
\includegraphics[width=0.23\textwidth]{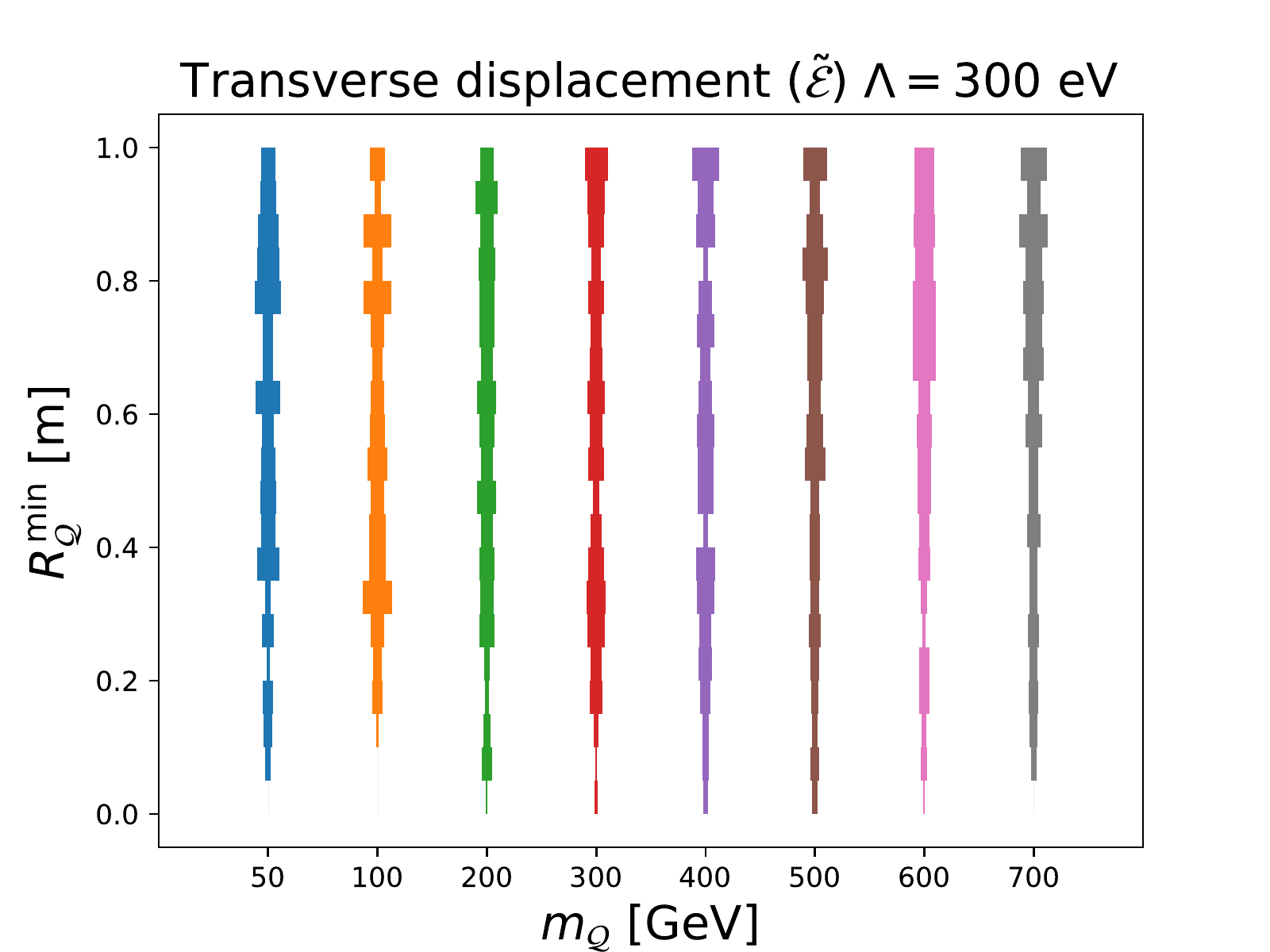}
\includegraphics[width=0.23\textwidth]{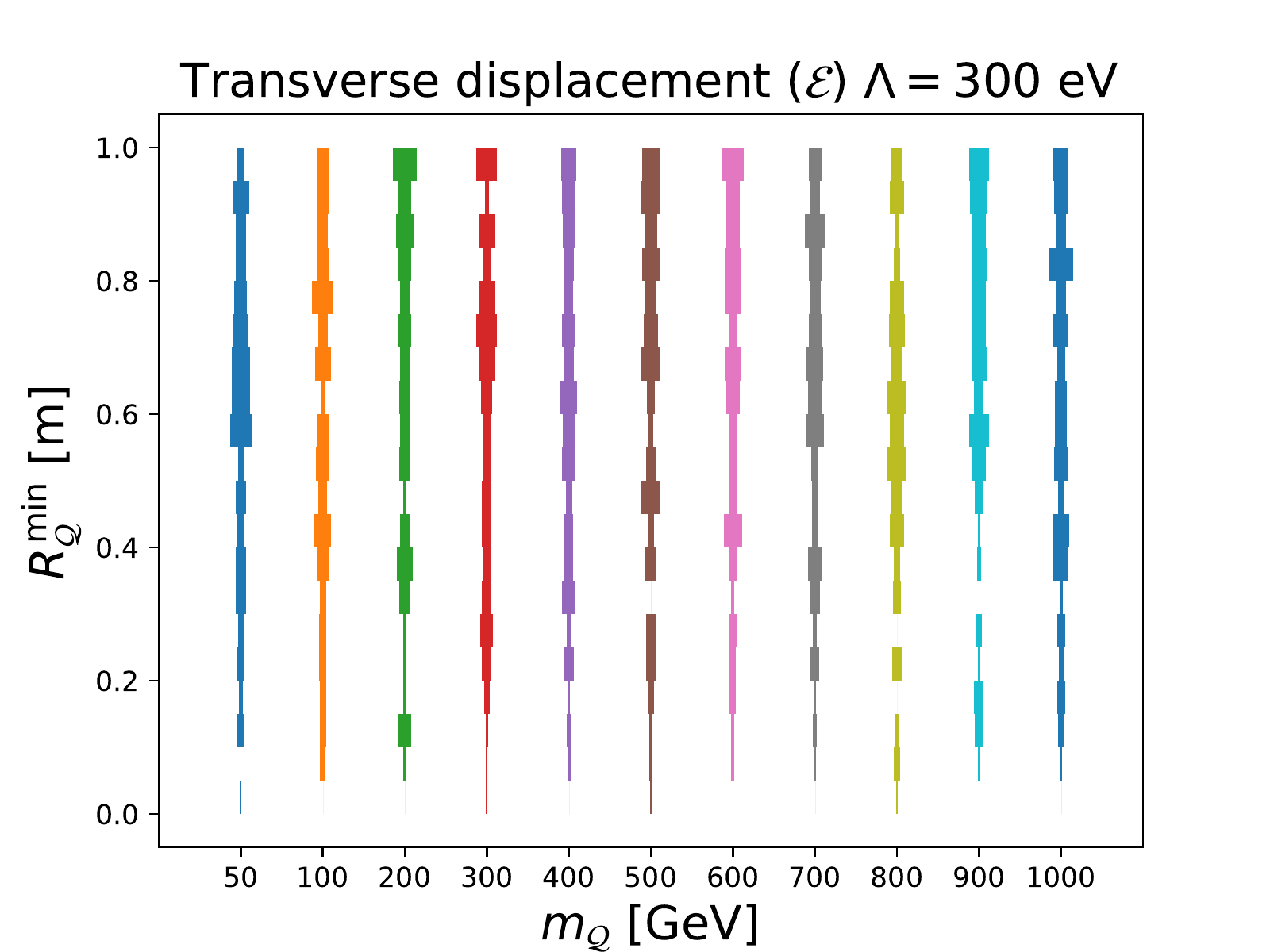} 
\includegraphics[width=0.23\textwidth]{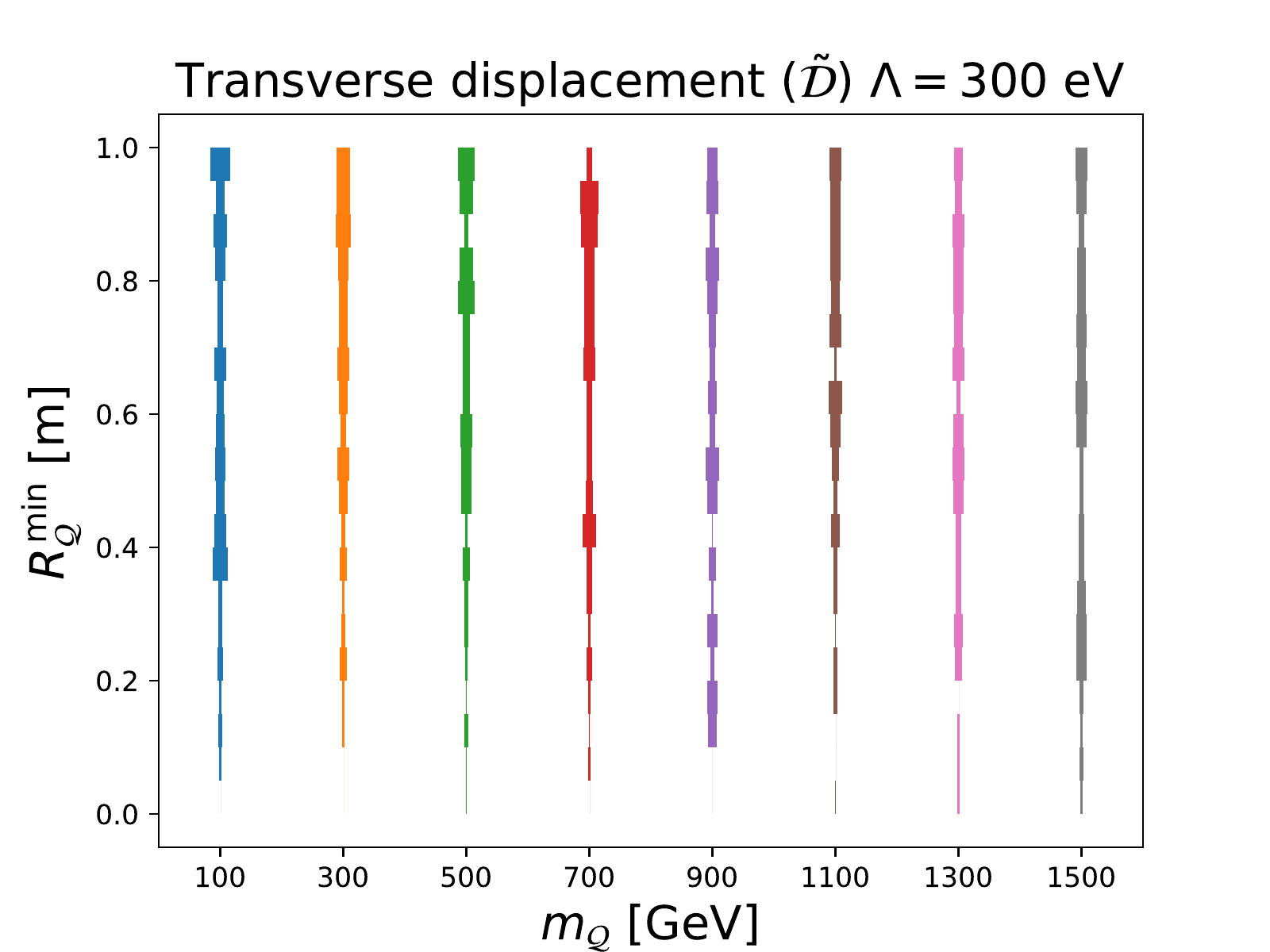}
\includegraphics[width=0.23\textwidth]{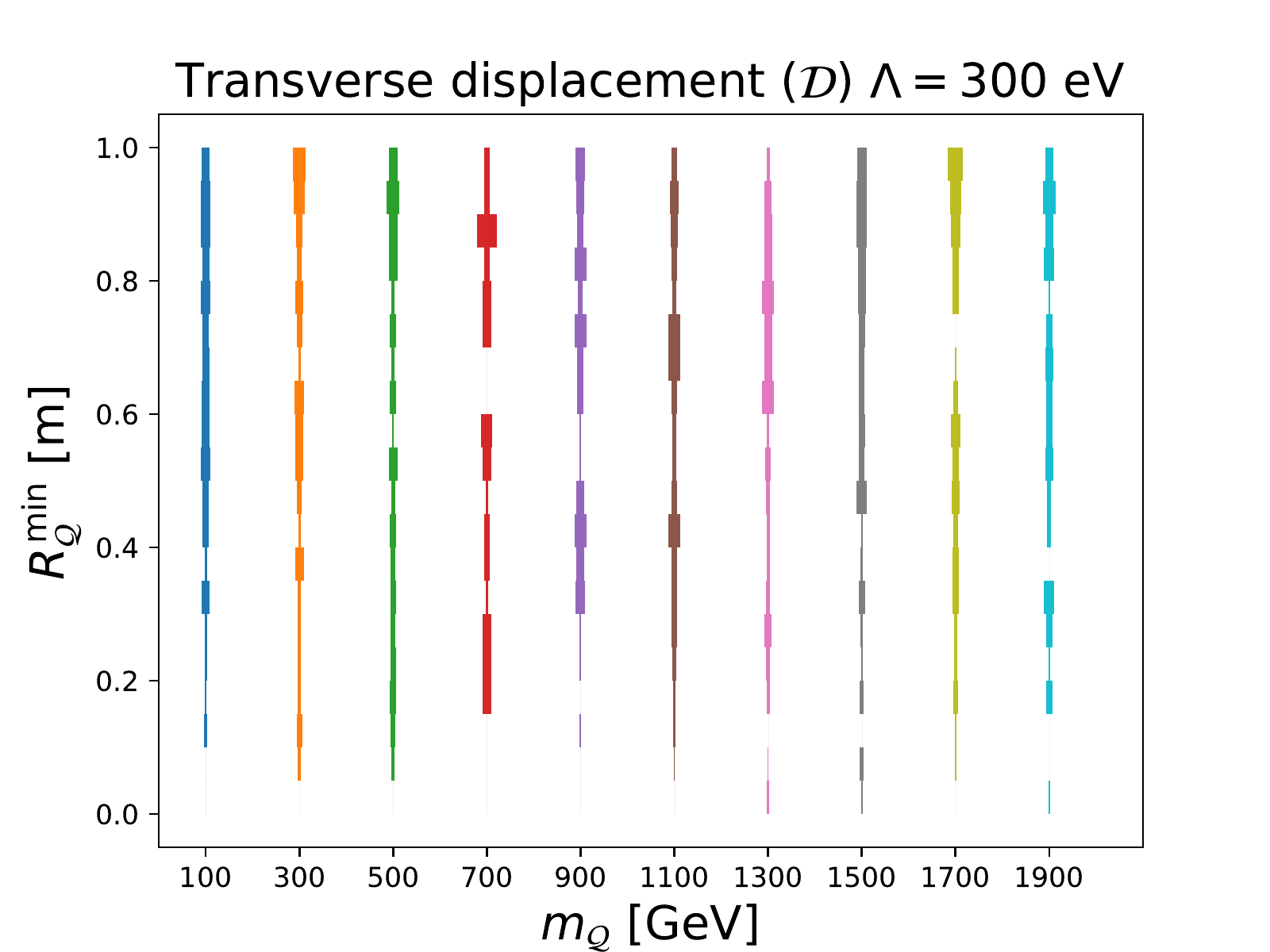} 
		\end{center}
		\caption{The distributions of the minimal transverse displacement ($R_{\mathcal{Q}}^{\min}$) of quirk hits in the FASER 2 tracker for different quirk masses and fixed $\Lambda$. The width of the band indicates the event number in a bin of $R_{\mathcal{Q}}^{\min}$. Upper panels: $\Lambda =100$ eV; Lower panels: $\Lambda=300$ eV. }
		\label{fig::trans13}
	\end{figure}
	
	As has been discussed in previous section, the position where the quirk enters the FASER (2) tracker is highly correlated with the quirk oscillation amplitude and also mildly depends on the quirk quantum numbers (because quirks with different quantum numbers have different momenta distributions in their productions). 
	In Figure~\ref{fig::trans13}, we present the distributions of the minimal transverse displacement ($R_{\mathcal{Q}}^{\min}$) of quirk hits in the FASER 2 tracker for different quirk masses and fixed $\Lambda$. 
	Events with at least one quirk entering the FASER 2 tracker are counted. For event with both quirks reaching the FASER 2 tracker, the one with smaller transverse displacement ($R_{\mathcal{Q}} = \sqrt{x^2_{\mathcal{Q}} + y^2_{\mathcal{Q}}} $) is chosen. 
	For $\Lambda \lesssim 100$ GeV,  many quirks from events with $p_T(\mathcal{QQ})/|p(\mathcal{QQ})| \gtrsim 0.002$ will enter the FASER 2 tracker due to the $\mathcal{O}(1)~\text{m}$ oscillation amplitude.
	For $\Lambda \gtrsim 400$ GeV (corresponds to $L \lesssim \mathcal{O}(10)~\text{cm}$), in most cases, only events with $p_T(\mathcal{QQ})/|p(\mathcal{QQ})| < 0.002$ can reach the FASER 2 tracker. The distributions of $R_{\mathcal{Q}}^{\min}$ spread over the range of $[0,1]$ m. 
	
	For events with both quirks entering the FASER 2 tracker, the distributions of the distance between the two hits on one tracking plane ($\Delta l_{\mathcal{Q}}$) as well as the distributions of the time difference between them ($\Delta t_{\mathcal{Q}}$) can be used to resolve the underlying quirk model parameters. 
	Although the number of events that can be used to calculate the $\Delta l_{\mathcal{Q}}$ and $\Delta t_{\mathcal{Q}}$ is smaller than that for calculating the $R_{\mathcal{Q}}^{\min}$. The distribution of $\Delta l_{\mathcal{Q}}$ contains more refined information about the quirk oscillation amplitude than that of $R_{\mathcal{Q}}^{\min}$. 
	Combining the information of $\Delta l_{\mathcal{Q}}$ and $\Delta t_{\mathcal{Q}}$, one can also estimate the oscillation period as given in Eq.~\ref{eq::tp}. 
	After obtaining the oscillation amplitude and oscillation period for the quirk signal, one can infer the confinement scale, quirk mass as well as the quantum numbers of the quirk.

\section{Conclusion}\label{sec5}

The quirk EoM is solved in a way that the Lorentz force and the radiations of force mediator are ignored. The ionization effects in different materials are treated carefully. We include all the important forward infrastructures from the ATLAS IP to the FASER (2) detector, whose effects on the quirk trajectory can not be ignored. The Gaussian distribution of the ionization energy loss for a charged quirk in the BB region is also taken into account.

The ionization forces on quirks will induce angular momentum of the quirk-pair system. The analytical calculations show that in the CoM frame the torques on the two quirks from a pair cancel with each other in a complete oscillation and thus have no contribution to the angular momentum. This indicates that the angular momentum of the quirk pair only changes sharply when the quirk pair injects into or moves out of the material, or just one quirk from the pair is travelling in the material.  

We take FASER 2 as an example to demonstrate the quirk signal efficiency in the detector tracker. Due to the lower velocity, heavier quirk suffers from stronger ionization force, which will change the moving direction of the quirk pair. What is more important, heavier quirk mass leads to a larger oscillation amplitude of the quirk pair, and thus makes the quirk pair more likely to bypass the FASER 2 tracker when  $p_T(\mathcal{Q}\mathcal{Q})/ |p(\mathcal{Q}\mathcal{Q})|$ is fixed. So, the signal efficiency becomes lower for heavier quirk.
When the confinement scale $\Lambda$ is sizable such that the characteristic oscillation amplitude $L \ll \mathcal{O}(1)~\text{m}$, only the quirk pairs satisfying $p_T(\mathcal{QQ})/|p(\mathcal{QQ})|<0.002$ can enter the FASER 2 tracker. 
The signal efficiency is very low in the small $\Lambda$ region where $L \gtrsim \mathcal{O}(10)~\text{m}$, because many quirk events will bypass the FASER 2 tracker even though the $p_T(\mathcal{QQ})/|p(\mathcal{QQ})|<0.002$ condition is fulfilled.
The signal efficiency is highest in the moderate $\Lambda$ region where $L\sim\mathcal{O}(1)~\text{m}$. In this region, beside the events with $p_T(\mathcal{QQ})/|p(\mathcal{QQ})|<0.002$, others with $p_T(\mathcal{QQ})/|p(\mathcal{QQ})| \gtrsim 0.002$ will also enter the FASER 2 tracker due to the sizable oscillation amplitudes. 

The total energy losses due to infracolor glueball and electromagnetic radiations can be estimated, whose values are both much smaller than the typical kinetic energy of quirk at the LHC. So, we ignore the effects of them in our simulation.

For an integrated luminosity of 3000 (150) fb$^{-1}$, given negligible background, FASER 2 (FASER) will be able to exclude the $\mathcal{E}$, $\mathcal{D}$, $\tilde{\mathcal{E}}$ and $\tilde{\mathcal{D}}$ quirks with mass below 990 (360) GeV, 1800 (900) GeV, 630 (200) GeV and 1280 (570) GeV, respectively, when $\Lambda \gtrsim \mathcal{O}(100)$ eV. Compared with the HSCP search, the mono-jet search, and the coplanar search for fermionic quirks, FASER 2 is much more sensitive than other searches when $\Lambda \gtrsim \mathcal{O}(100)$ eV. For the color neutral quirk $\mathcal{E}$, FASER behaves better than other searches when $\Lambda \gtrsim 150$ eV.

There are some other features of the quirk signal can be used to resolve the model parameters. For events with at least one of the two quirks entering the FASER (2) tracker, the distributions of the minimal transverse displacement ($R_{\mathcal{Q}}^{\min}$) for quirk hits are closely related to the confinement scale $\Lambda$ of the new gauge group. For event with both quirks reaching the FASER (2) tracker, the distributions of the distance between the two quirk hits on one tracking plane ($\Delta l_{\mathcal{Q}}$) as well as the distributions of the time difference between them ($\Delta t_{\mathcal{Q}}$) will also be very useful to resolve the model parameters.


\begin{acknowledgments}
This work was supported in part by the Fundamental Research Funds for the Central Universities, by the NSFC under grant No. 11905149, by the Projects 11875062 and 11947302 supported by the National Natural Science Foundation of China, and by the Key Research Program of Frontier Science, CAS.

\end{acknowledgments}

\bibliographystyle{jhep}
\bibliography{quirk_faser}
\end{document}